\documentclass[a4paper, 11pt]{article}

\usepackage{microtype}
\usepackage{verbatim}
\usepackage[margin=1in]{geometry}
\usepackage[singlespacing]{setspace}
\usepackage{multirow}

\usepackage{xr}

\usepackage{hyperref}
\usepackage{url}
\usepackage{amsmath}
\usepackage{amssymb}

\usepackage{dsfont}
\usepackage{upgreek}
\usepackage{bbm}
		
\usepackage{graphics}
\usepackage{graphicx}
\usepackage{epstopdf}
\usepackage{tikz}
\usepackage{float}

\usepackage{booktabs}
\usepackage{subfigure}

\usepackage[square, sort, numbers]{natbib}
\usepackage{xspace}

\usepackage[scaled=.7]{beramono}

\usepackage{trimclip}

\usepackage{algorithm}
\usepackage{algorithmicx,algpseudocode}
\usepackage{enumitem}

\usepackage{amsthm}
\newtheorem{theorem}{Theorem}
\newtheorem{lemma}{Lemma}

\newtheorem{proposition}{Proposition}
\newtheorem{definition}{Definition}
\newtheorem{remark}{Remark}
\newtheorem{assumption}{Assumption}

\newcommand{\DMfull}{decision maker\xspace}
\newcommand{\DM}{DM\xspace}

\newcommand{\SP}{secretary problem\xspace}

\newcommand{\SSPfull}{Sequential Selection Problem\xspace}
\newcommand{\SSP}{SSP\xspace}
\newcommand{\SSPs}{SSPs\xspace}

\newcommand{\WSSPfull}{Warm-starting SSP\xspace}
\newcommand{\WSSP}{WSSP\xspace}
\newcommand{\ISSP}{GSSP\xspace}
\newcommand{\ISSPs}{GSSPs\xspace}
\newcommand{\ISSPfull} {Generalized SSP\xspace}

\newcommand{\MSSPfull}		 	{Multi-round Sequential Selection Problem\xspace}
\newcommand{\MSSP}		 			{MSSP\xspace}

\newcommand{\background} 		{background\xspace}
\newcommand{\Background} 		{Background\xspace}
\newcommand{\bckrnd} 				{\mathcal{B}}
\newcommand{\sample}    		{\sample\xspace}
\newcommand{\candidate}    	{candidate\xspace}
\newcommand{\candidates}    {candidates\xspace}
\renewcommand{\sample}    	{sample\xspace}

\newcommand{\algofull}    	{Cutoff-based Cost Minimization\xspace}
\newcommand{\algo}    	 		{CCM\xspace}
\newcommand{\algoNewfull}	  {low failures-\algo}
\newcommand{\algoNew}			  {lf-\algo}
\newcommand{\budget}    		{b}
\newcommand{\quality}    		{q}
\newcommand{\policy}    		{\pi}
\newcommand{\nres}					{r} 
 
\newcommand{\pres}		  		{\Prob_{r}}
\newcommand{\An}		        {\tilde{A}_n}
\newcommand{\an}		        {\tilde{A}_n}
\newcommand{\ajm}		        {\tilde{A}_{j-1}}
\newcommand{\aj}		        {\tilde{A}_{j}}

\newcommand{\captionSize}   {\footnotesize}
\newcommand{\B}				      {O}	 

\newcommand{\rating}				{rating\xspace}  

\newcommand{\score}					{score\xspace}  
\newcommand{\scores}				{scores\xspace}  

\newcommand{\Xbold}	        {\mathbf{X}} 
\newcommand{\val}			      {S}  
\newcommand{\Sbold}	        {\mathbf{\val}}  

\renewcommand{\S}[1]				{ \val_{#1} } 
\newcommand{\X}[1]					{ X_{#1} }

\newcommand{\Xrefbold}[1]		{ \mathbf{Y}_{#1} } 
\newcommand{\Xref}[2]				{Y_{(#1),#2}}

\newcommand{\symbolpres}		{dot\xspace}
\newcommand{\symb}	  			{\dot}
\newcommand{\Spres}					{\symb{\val}}  
\newcommand{\Xpres}					{\symb{X}}
\newcommand{\Apres}					{\symb{A}}

\newcommand{\Sobold}		  	{ \symb{\mathbf{\val}} } 
\newcommand{\So}[1]			 		{\Spres_{(#1)}}  
\newcommand{\Xobold}		    {\symb{\mathbf{X}}} 
\newcommand{\Xo}[1]		      {\Xpres_{(#1)}}   
	
\newcommand{\Xavailbold}	  { {\symb{\mathbf{X}}}^+ }
\newcommand{\Xavail}[1]		  {  {\Xpres}^{+}_{(#1)} }

\newcommand{\Abold}		      { \mathbf{A} }
\newcommand{\A}[1]					{A_{#1}}

\newcommand{\Aobold}[1]	    {\symb{\mathbf{A}}_{#1}}  
\newcommand{\Ao}[2]		      {\Apres_{(#1),#2}}  
\newcommand{\coff} 					{\rsymb^*\xspace} 

\newcommand{\xmin} 					{1}
\newcommand{\xmax} 					{n+b}
  
\newcommand{\smallo} 				{\sigma^2_{j-1}}
\newcommand{\smalloj}				{\sigma^2_{j}}
\newcommand{\pind}					{\frac{\gamma_i-1}{n+b} }
\newcommand{\pjnd}					{\frac{\gamma_j-1}{n+b} }

\newcommand{\rankset}[1] 		{\mathcal{P}_{#1}}
\newcommand{\oneoverb}			{ }  

\newcommand{\scalar} 				{^\top}   
\newcommand{\regret} 				{regret\xspace}  
\newcommand{\rsymb}    		 	{\phi}  
\newcommand{\referent}    	{referent\xspace}  
\newcommand{\referents}    	{referents\xspace} 
\newcommand{\refset}   			{reference set\xspace}
\newcommand{\refsets}  			{reference sets\xspace}
\newcommand{\updref}  			{updated \refset}
\newcommand{\rank} 					{R}
\newcommand{\loss} 					{loss\xspace}

\makeatletter
\g@addto@macro\bfseries{\boldmath}
\makeatother

\newcounter{phase}[algorithm]
\newlength{\phaserulewidth}
\newcommand{\setphaserulewidth}{\setlength{\phaserulewidth}}
\newcommand{\phase}[1]{%
  \vspace{-1.45ex}
  \Statex\leavevmode\llap{\rule{\dimexpr\labelwidth+\labelsep}{\phaserulewidth}}\rule{\linewidth}{\phaserulewidth}
  \Statex\strut\refstepcounter{phase}\!\!\!\!\!\!\!\!\raisebox{0.25em}{{\scriptsize$_\blacksquare$}}\ \textit{\textbf{#1}}
  \vspace{-1.25ex}\Statex\leavevmode\llap{\rule{\dimexpr\labelwidth+\labelsep}{\phaserulewidth}}\rule{\linewidth}{\phaserulewidth}}
\makeatother
\setphaserulewidth{.3pt}

\newcommand{\Sec}[1]		{Sec.\,\ref{#1}}
\newcommand{\Fig}[1]		{Fig.\,\ref{#1}}

\newcommand{\Eq}[1]			{Eq.\,\ref{#1}}

\newcommand{\Alg}[1]		{Alg.\,\ref{#1}}
\newcommand{\Theorem}[1]{Theorem~\ref{#1}}

\newcommand{\Proposition}[1]{Proposition~\ref{#1}}

\newcommand{\Lemma}[1]{Lemma~\ref{#1}}
\newcommand{\Definition}[1]{Definition~\ref{#1}}

\newcommand{\ie}   			{i.e.\@\xspace}
\newcommand{\eg}   			{e.g.\@\xspace}
\newcommand{\etc}   		{etc.\xspace}
\newcommand{\wrt}   		{w.r.t.\@\xspace}
\newcommand{\iid}   		{i.i.d.\@\xspace}
\newcommand{\st}   			{s.t.\@\xspace}

\newcommand{\one}       {\mathbf{1}}
\newcommand{\oneS}[1]   {\one_{[#1]}}

\newcommand{\ind}       {\mathds{1}}
\newcommand{\Ind}[1]    { \ind{\{#1\}} }

\newcommand{\Exp}[1]    {\mathbb{E}[#1]}
\newcommand{\real}      {\mathbb{R}}
\newcommand{\Prob}      {\mathbb{P}}

\newcommand{\mydots} 	{...}
\newcommand{\algComment}[1] 	{\hfill/\!/\,{#1}}

\newcommand{\INTerval}[2] 	{\{#1,\mydots, #2\}}

\newcommand{\inlinetitle}[2]  {\vspace{4pt}\noindent\textbf{\emph{#1}{#2}}}

\renewcommand*{\top}{{\mkern-1.5mu\mathsf{T}}}

\newcommand{\mySqBullet}		{\raisebox{0.25em}{{\scriptsize$_\blacksquare$}}}

\renewcommand{\t}				{j}		
	
\renewcommand{\Prob}			{\textup{P}}	

\newcommand\blankfootnote[1]{%
  \begingroup
  \renewcommand\thefootnote{}\footnote{#1}%
  \addtocounter{footnote}{-1}%
  \endgroup
}

\begin{document}

\title{The Warm-starting Sequential Selection Problem \\and its Multi-round Extension}

\author{Mathilde Fekom\,\quad Nicolas Vayatis\,\quad Argyris Kalogeratos\blankfootnote{The authors are with the Center of Applied Mathematics (CMLA) -- ENS Paris-Saclay, University Paris-Saclay, 94230 Cachan, France. Contact emails: \texttt{\footnotesize \{fekom,\,vayatis,\,kalogeratos\}@cmla.ens-cachan.fr}.\qquad\qquad\qquad\qquad\qquad\qquad Part of this work was funded by the IdAML Chair hosted at ENS Paris-Saclay.}}

\date{}

\maketitle

\begin{abstract}
In the \emph{\SSPfull} (\SSP), immediate and irrevocable decisions need to be made as \candidates randomly arrive for a job interview. 
Standard SSP variants, such as the well-known \SP, begin with an empty selection set (\emph{cold-start}) and perform the selection process once over a single \candidate set (\emph{single-round}).
In this paper we address these two limitations. First, we introduce the novel \emph{\WSSPfull} (\WSSP) setting which considers at hand a \emph{\refset}, a set of previously selected items of a given \emph{quality}, and tries to update optimally that set by (re-)assigning each job at most once. 
We adopt a \emph{cutoff-based} approach to optimize a rank-based objective function over the final assignment of the jobs. 
In our technical contribution, we provide analytical results regarding the proposed \WSSP setting, we introduce the algorithm \emph{\algofull} (\algo) (and the \emph{\algoNewfull}, which is more robust to high rate of resignations) that adapts to changes in the quality of the reference set thanks to the \emph{translation method} we propose. Finally, we implement and test \algo in a multi-round setting that is particularly interesting for real-world application scenarios. 
\end{abstract}
\maketitle
\newpage

\section{Introduction}\label{sec:introduction}

Since its introduction in the early 60's, the \SP \cite{Lindley61, Freeman83, Ferguson89} has been perhaps the most famous optimal stopping problem: $n$ randomly incoming \candidate secretaries are interviewed one after the other for a job position. In each interview, the \emph{\DMfull}  (\DM) acquires information about a \candidate's competence which allows her to rank him among the so far examined \candidates. She can decide when to terminate the process by selecting the last \candidate interviewed. 
The \DM has no knowledge of who will come later on, yet her decisions should be \emph{immediate} and \emph{irrevocable} after each interview. This describes a \emph{\SSPfull} (\SSP\footnote{Depending on the context, the last letter of the abbreviations \SSP and the herein presented \MSSP may refer to the respective selection `\emph{Problems}' or the associated selection `\emph{Processes}'.}). The class of \SSP problems is attractive for theoretical analysis and for practical use, due to its generality and evident relevance to online selection under realistic constraints. 
Same as in this work, SSPs are usually presented in the intuitive recruitment context.

The goal of the original problem is to select none but the best among the sequence of $n$ \candidates, while in each interview the \DM only realizes the relative quality of the examined \candidate, that is his relative rank.
The standard algorithm, first proposed in \cite{Lindley61}, is a \emph{cutoff-based} approach which comprises two phases: the \emph{learning phase} where a number (referred to as \emph{cutoff}) of \candidates are automatically rejected, and the \emph{selection phase} where the first \candidate ranked above the best recorded during the first phase is hired (or the last one, by default). In essence, the former phase learns a threshold that is subsequently used in the latter to spot the first \candidate to beat it. 
For instance, the optimal cutoff for maximizing the probability to find the best \candidate is $c^*= \lfloor n/e \rfloor$ asymptotically. 
Note that the multi-choice problem is a natural extension of the above (see \Sec{sec:related_work}).

\inlinetitle{Motivation and contribution}{.} %
Our motivation derives from real-world recruitment processes that take place in large organizations or companies whose aim is to dynamically adapt in their operating environments. 
This setting goes beyond the existing \SSP models in literature that have one important limitation, namely they consider a \emph{cold-start} initialization where there is no assignment of jobs at the beginning of the selection process.

To address this issue, we introduce a new online \emph{initialized}  problem that we call \emph{\WSSPfull} (\WSSP): at the beginning of the selection, the \DM has at hand a \emph{\refset} of \referents for whom she knows the status of \emph{availability} (\referents are allowed to quit their jobs just before the beginning of the interviews), and eventually the \emph{average quality} \wrt the new \candidates. 
The selection strategy operates as in the standard cutoff-based fashion, however having a \refset of a given size, the question whether the learning phase of the \DM should be longer or shorter is not obvious.
We thereby propose an algorithm for the \WSSPfull, called \emph{\algofull} (\algo), that gives the optimal learning time (\ie the optimal cutoff value) according to the main parameters of the problem, while trying to minimize a \emph{\regret} defined as the average sum of the ranks of the selected items. 

As for the technical contributions, we analyze the \WSSPfull and derive analytical formulas for: i)~the initialization, specifically the expected rank of the \referents (available or not) and the minimal \regret of an offline strategy, and ii)~the expectation of the main parameters of the process when using \algo, \ie the acceptance threshold for each candidate, the number of new hires, and the \regret. From the latter, we infer the optimal cutoff $c^*(\budget,\nres,n)$, given the number of jobs $\budget$, the number of \candidates $n$, and the number of resignations $\nres$ starting with the case where the quality of the \refset is average; thereafter we propose a translation method that permits to derive $c^*$ for every value of the quality $\quality$ and highlights some interesting results. We then propose the \emph{\algoNewfull} (\algoNew) variation that is more robust to high resignation rates and hence prevents from accepting the very last \candidates by default.  

The rest of the paper is organized as follows. \Sec{sec:related_work} presents the background of our work including related research; in \Sec{sec:ssp} we present a new formalism for a broad range of \SSPs called \ISSPfull (\ISSP), and introduce one of its specific instance, the \WSSPfull. \Sec{sec:algo} details the proposed \algo algorithm, tries to answer the question of the optimal learning time, describes the \emph{translation method} and the \algoNew. Then, \Sec{sec:mssp} gives an implementation of the \algo algorithm in a multi-round fashion and, finally, our conclusions and future work are presented in \Sec{sec:conclusions}.

\section{Related work} \label{sec:related_work}

Various extensions of the basic \SP have been investigated; for non-exhaustive surveys see \cite{Freeman83, Ferguson89}. Importantly, a change in the setting or in the objective function, changes also the optimal cutoff. In some scenarios, the \DM can not only compute the \emph{relative rank} of an interviewed \candidate among those examined earlier, but also assess \candidate's true \emph{quality \score}. 
This \score can be thought of as a random variable associated with each \candidate. 
In \cite{Bearden06}, \candidates are drawn from a uniform distribution on [0,1] but  the \DM can only rank \candidates relatively to those she has seen before, and the objective is to maximize the expectation of the \score of the selected \candidate. They have shown that in this case, the optimal cutoff becomes $c^*=\sqrt{n}-1$. On the other end, \emph{Robbin's problem} \cite{Bruss05} 
seeks to minimize the expectation of the rank of the selected \candidate (note: low ranks are better). However, the analytical solution to this problem remains unknown, even when the \score distribution of the \candidates is known. 

Notable variants are those related to \emph{multiple stopping}, or simply $\budget$-choice, where the \DM has to select $\budget$ \candidates \cite{Kleinberg05, Mosteller66, Bearden06bis, MBateni13, Krieger10, Babaioff07, Broder09, Nikolaev07}. In that case, the objective set function can be modular (\ie equivalent to adding up the independent application of the function to the set of elements), submodular \cite{MBateni13, Feldman17}, or subject to matroid constraints \cite{Feldman15, Feldman18, Soto18}. Non-modularity introduces interesting set evaluation aspects, such as the complementarity or mutual-enhancement among the selected \candidates, which are however out of the scope of this work. Regarding modular objective functions, \cite{Babaioff07} studies the $\budget$-choice problem with the objective to maximize the sum of \scores of the selected \candidates, that arrive in a random order, without assuming prior knowledge of the \score distribution. An interesting finding is that the optimal cutoff for that setting does not depend on $\budget$: $c^* = \lfloor n/e \rfloor$. 

Very few papers study the algorithmic notions related to repeated selections \cite{Stewart78}, as well as the human capacity to learn the right cutoff after reviewing multiple independent \candidate sets \cite{Goldstein17, Bearden06bis}. However, \cite{Stewart78} develops a non cutoff-based strategy which is implemented regarding two distinct aims: to maximize the probability of selecting the best, or to maximize the expected \score of the selected \candidate. That work concludes by stating that learning the \score distribution does contribute to the efficiency of the selection only \wrt the second aim. 
An experimental comparison of simpler and intuitive non cutoff-based heuristics is provided in \cite{Seale97}. More sophisticated adaptive strategies worth to be mentioned are the \emph{Bruss' odds theorem} \cite{Bruss00} and the work in \cite{Nikolaev07}. A rather different scenario concerns a startup company (or a new ambitious business unit) which is initially funded by a handful of people but is about to grow larger. The so-called \emph{hiring problem} \cite{Broder09} refers to the \SSP that aims at driving the optimal growth of personnel using an adaptive selection threshold based on the already employed items. Among heuristics, such as hiring above the worst or the best current \referents, hiring above the mean \referent \score shown to be the best performing strategy. Similar settings where a set of selected candidates increases through time are considered in \cite{Krieger07, Krieger08, Helmi14, Fiat15}, while \cite{Helmi14} makes a thorough analysis of \emph{hiring above the $m$-th best} strategies. In \cite{Fiat15} the \emph{temp \SP} is introduced where contracts are of a fixed duration, thus temporary. The improved algorithm presented in \cite{Kesselheim16} generalizes towards general packing constraints and arbitrary hiring durations. 

\section{A general class of Sequential Selection Processes and the novel Warm-starting setting}

\inlinetitle{Notations}{.}
A bold symbol denotes a vector, for instance, $\mathbf{A} = (A_1,\mydots,A_k) \in \real^k$, $\forall k \in \mathbb{N}^*$, in which with little abuse we omit the symbol of the transpose. The concatenation of matrices is denoted by $(\mathbf{A},\mathbf{B})$. Moreover, $\Ind{\cdot}$ is the indicator function, which is $1$ if the input condition is true, and otherwise $0$; also, $\oneS{l}$ is the unit vector of length $l$. 

\subsection{Generalized Sequential Selection Process}

In a standard \emph{Sequential Selection Process} (\SSP), \emph{candidates} for a job position arrive sequentially in random order. The qualitative skills of each candidate can be assessed independently on his arrival by the \emph{\DMfull} (\DM), allowing the relative ranking of the examined candidates against each other. According to this evaluation, the \DM chooses who to hire in order to optimize a given \emph{objective function}. 
\begin{definition}{\ISSPfull (\ISSP):} Online selection process described by the following elements organized in several categories:
\begin{enumerate}[itemsep=-2pt,leftmargin=*]
\item \Background
\noindent $\bckrnd$: collection of information known upfront by the \DM, including the set $\mathcal{A}$ of all possible actions the \DM can take (\eg hire, fire, add in queue, put on standby, \etc).
\item Sequential Arrivals 
\vspace{-2mm}
\begin{itemize}[label=--,itemsep=0pt,leftmargin=*]
\item $\Sbold = (\S{j})_{j\ge 1}$: sequence of candidate scores \st $\S{j} \in \mathcal{S} \subset \real$, drawn from distribution $f_j, \, \ \forall j$.
\end{itemize}
\item Decision Process
\begin{itemize}[label=--,itemsep=0pt,leftmargin=*]
\vspace{-2mm}
\item $\policy =(\policy_j)_{j\le1}$: policy, \ie sequence of mappings where $\policy_j : \mathcal{S}^j \times \mathcal{A}^{j-1} \rightarrow \mathcal{A}$;
\item $\Abold = (\A{j})_{j\geq 1}:$ sequence of decisions regarding the candidates, according to the policy, \ie  $\A{j}= \pi_j(\S{1},...,\S{j},\A{1},...,\A{j-1}) \in \mathcal{A}, \, \ \forall j$. 
\end{itemize}
\item Evaluation
\vspace{-2mm}
\begin{itemize}[label=--,itemsep=0pt,leftmargin=*]
\item$\ell: \mathcal{S} \times \mathcal{A} \rightarrow \real_+$: loss function \st $\ell(s,a)$ is the \loss for taking decision $a$ after observing $s$;
\item $L(\Sbold,\Abold) = \sum_{j\le1} \ell(S_j, A_j) $: cumulative loss;
\item  Let $P$ be the distribution of $(\Sbold,\Abold)$. The evaluation criterion, called \emph{regret}, is evaluated at the end of the process and defined as $\Phi(\policy) = \mathbb{E}_P[\rsymb_\bckrnd(\Abold\mid{\Sbold}) ]$, where:
\vspace{-1mm}
\begin{equation}
\rsymb_\bckrnd(\Abold\mid \Sbold) =\,\, \mid L(\Sbold,\Abold) -\rsymb^*\mid  \,\, \in \real_+,
\label{eq:cost_general}
\end{equation}
and $\rsymb^*$ is a baseline value.
\end{itemize}
\end{enumerate}
\label{def:GSSP}
\end{definition}
With the high-level formalization of the \ISSP class, we can summarize several well-known processes, such as the indicative ones mentioned below.

\inlinetitle{Examples of well-known \ISSPs}{:}
\vspace{0.3em}

\noindent\mySqBullet~Standard \SP \cite{Dynkin63}: A \ISSP setting where $\bckrnd = (b, n, \mathcal{A} = \{0,1\})$, where $b=1$ is the number of job position, $n$ is the finite number of candidates.
and a candidate is either selected (hired, $\A{j}=1$) or rejected ($\A{j}=0$). It is assumed that decisions are immediate and irrevocable, that candidates arrive in a random order, and that their scores are not independent (each candidate's score depends on those examined before) nor identically distributed. This is equivalent to having relative ranks as observations, \ie a triangular array $(X_{i,j})$ where $X_{i,j} \in \{1,...,j\}$ is the relative rank of the $i$-th incoming candidate after having examined $j \ge i$ of them. The vector of absolute ranks, evaluated at the end, is given by $\Xbold = ( X_{1,n},...,X_{n,n}) \in \rankset{n}$, where $\rankset{l}$ is the set of all permutations of the elements of $\{1,...,n\}$.
 The evaluation criterion to maximize is the probability to select the best candidate (the one with absolute rank 1 at the end of the selection), which can be expressed by $\ell(X_{j,j},\A{j}) = \Ind{X_{j,n}\A{j}=1}, \, \ \forall j \le n$, therefore $\rsymb^*=0$ and $\Phi = \Prob(\Xbold\scalar \Abold =1)$.
\vspace{0.4em}

\noindent\mySqBullet~Hiring problem \cite{Broder09}: A multi-choice \ISSP setting where, by respecting the trade-off between the rate of hires and the quality of the hired candidates, the objective is to grow the company as much as possible while keeping maximal the average \score of the employees. The recruitment process has infinite horizon. Therefore we have $\bckrnd = (b, n, \mathcal{A} = \{0,1\})$, where $b\rightarrow \infty$ and $n\rightarrow \infty$. It is assumed that decisions are immediate and irrevocable, and that observations are \iid scores drawn from a uniform distribution, \ie $S_{j} \sim \mathcal{U}(0,1)$.

\begin{remark} The number of job positions $b$ and of candidates $n$ are usually included in the  \background $\bckrnd$; however variants of the standard \SP \cite{abdel-hamid82, Krasnosielska-Kobos15} may involve a random number of candidates. Note that $b>1$ usually refers to a {multi-choice} or {multi-stopping} problem.
\end{remark}

\begin{remark}In most \ISSP settings, the \loss suffered at each decision is the score of an accepted item, \ie $\ell(\S{j},\A{j})=\pm\S{j}\A{j}$, with a positive (resp. negative) sign if the goal is to minimize (resp. maximize) the sum of scores. The evaluation criterion is further detailed into two cases: 1) the `{no regret}' case, where the \DM merely tries to optimize its selection \ie for $\rsymb^*=0$, and 2) the `{with regret}' case, where the online selection is to be compared to the best associated offline selection $\policy_{\text{off}}$, where the \DM knows the entire sequence of candidates beforehand, in this case  $\rsymb^* = \underset{\mathbf{\B} \in \policy_{\text{off}}}{\min} \rsymb(\Sbold,\mathbf{\B})$ (or $\rsymb^* = \underset{\mathbf{\B} \in \policy_{\text{off}}}{\max} \rsymb(\Sbold,\mathbf{\B})$ when the goal is to maximize the sum of the scores).
\end{remark}

\subsection{The Warm-starting Sequential Selection Process}\label{sec:ssp}

\inlinetitle{Description and rules of the game}{.}
The \emph{\WSSPfull} (\WSSP) is a particular \ISSP instance that overcomes the limitations of standard cold-starting \SSP frameworks. 
Its characteristics is to start with a set of items at hand, called \emph{\refset} and composed of \emph{\referents}, each of them having also a status of \emph{availability}. The total number of job positions determines the size of the \refset. Items can therefore be of two types, \candidate or \referent. 
The value of each item is observed through a fixed real-valued \emph{relative score}, \ie each item's score depends on the scores of those already seen. Although the \referent's availability status can be broad (\eg on vacation, sick leave, resigned, \etc), we only allow resignations, \ie a \referent is \emph{unavailable} if he resigned (leaving his position empty) and \emph{available} otherwise (in other words, he is preselected).
In this paper, we work under the simple assumption that resignations are independent. The \DM therefore seeks highly-skilled candidates to 1)~fill up empty positions and 2)~replace non-competitive available \referents; by respecting the following specific constraints.
\begin{assumption} On the sequence of arriving candidates:

1.A)~Candidates arrive in a random order.

1.B)~Scores are not observed, the \DM can only make pairwise comparisons between items.
\end{assumption} 
\begin{assumption} On the decision policy:

2.A)~{The availability status is known upfront, and fixed throughout the process.}

2.B)~{Decisions are immediate and irrevocable.}

2.C)~Every position must be filled at the end of the process 

\end{assumption}

\inlinetitle{Formal definition}{.}
We add a \symbolpres on top of a variable to refer explicitly to the \refset, \eg $\Sobold = (\So{1},\So{2},...,\So{b}) \in \real^b$ gives the value represented by the variable $\Sbold$ (here, \scores) of the \referents in descending order: the best, the second best, \etc
Let the ranking function $\rank_{N} : \real \times \real^N \rightarrow \{1,...,N\}$, be the function that gives to each element of a collection of values its rank from 1 to $N$ when compared to the other values, \st $\rank_N(s, \mathbf{\Sigma}) = \sum_{i=1}^N \Ind{\Sigma_i \le s},\, \ \forall s \in \mathbf{\Sigma}$, where $\Sigma$ is a finite number set.
\\

\begin{definition}{\WSSPfull\!\! (\WSSP):} A particular \ISSP with the following characteristics:
\begin{enumerate}[itemsep=-2pt,leftmargin=*]
\item \Background  
\vspace{-1mm}

\noindent $\bckrnd =(n,b,\mathcal{A},\Aobold{0})$, where the included elements are:
\vspace{-2mm}
\begin{itemize}[label=-- ,itemsep=-2pt,leftmargin=*]
\item $n \in \mathbb{N}^*$: finite number of \candidates to appear;
\item $b \in \mathbb{N}^*$: number of job positions \st $b\le n$;  
\item $\mathcal{A} = \{0,1\}$: the set of possible actions the \DM can take, respectively reject or hire;
\item $\Aobold{0} = (\Ao{1}{0},\mydots,\Ao{b}{0}) \in \{0,1\}^{b}$: availability status of the \refset \st $\Ao{i}{0} = 1$ if the $i$-th best \referent is available.
\end{itemize}

\item Sequential Arrivals and  3.\,\,\,Decision Process {as in \Definition{def:GSSP}} 
\setcounter{enumi}{3}

\item Rank-based evaluation

\noindent The following simplified notation for the absolute ranks is written $\rank(s) = \rank_{b+n}(s,(\Sobold,\Sbold))$, where $\Sobold = (\So{1},\mydots,\So{b}) \in \real^b$ gives the \referents scores (sorted in descending value order for convenience). 
\begin{itemize}[label=--,itemsep=0pt,leftmargin=*]
\vspace{-1mm}
\item $\Xobold = \big(\Xo{1}=\rank(\So{1}) ,\mydots,\Xo{b} = \rank(\So{b}) \big) \in \real^{b}$: \referents' absolute ranks, 
\item $\Xbold = \big(\X{1}= \rank(\S{1}),\mydots,\X{n} = \rank(\S{n}) \big) \in \real^{n}$: candidates' absolute ranks, 

\item Let $P$ be the distribution of $(\Xbold,\Abold)$. The evaluation criterion, called \emph{regret}, is evaluated at the end of the process and defined as $\Phi(\policy) = \mathbb{E}_P[\rsymb_\bckrnd(\Abold\mid{\Xbold}) ]$, where:

\begin{equation}
\rsymb_\bckrnd(\Abold\mid{\Xbold}) = \left(\Xobold\scalar \Aobold{n} + \Xbold\scalar \Abold \right)\,\,\,\,\, -\underset{(\symb{\mathbf{\B}}_n, \,\mathbf{\B}) \in \policy_{\text{off},\bckrnd}}{\min} \left(\Xobold\scalar \symb{\mathbf{\B}}_n + \Xbold\scalar \mathbf{\B} \right)\, \ \in \real_+,
\label{eq:cost}
\end{equation}
where $\policy_{\text{off},\bckrnd}= \left\{ (\symb{\mathbf{\B}}_n, \,\mathbf{\B}) \in \{0,1\}^{n+b} :\,  (\symb{\mathbf{\B}}_n,\mathbf{\B})\scalar \oneS{n+b} = b\right\}$ and $\Aobold{n} \in \{0,1\}^b$ is the hiring decisions of the \referents after $n$ interviews of candidates.
\end{itemize}
\end{enumerate}
\label{def:wssp}
\end{definition}
The first term in \Eq{eq:cost} is the sum of the ranks of the items to which jobs have been assigned at the end of the selection. The second term is the minimal \regret achievable by an offline oracle strategy that, knowing the ranks, would select the best out of the available \referents (\ie for some $i$: $\Ao{i}{0} = 1)$ and the candidates. 
\begin{remark} 
In this work we make no assumptions at all about the source and nature of the scores. This is why we adopt a rank-based criterion to assess the selection strategy, which is a standard approach in nonparametric statistics.
\end{remark} 

\section{The proposed \algofull policy}\label{sec:algo}

In this section we present our novel algorithm for the \WSSP, called \emph{\algofull} (\algo). It takes as input a cutoff value $c\in \mathbb{N}$ representing the size of the \emph{learning phase}, \ie the number of candidates to be rejected by default from which the \DM learns valuable information about the overall sample. In the next section, we will analyze its optimality.

 \subsection{Cutoff-based strategies}

Inspired by the \SP, we develop the \emph{\algofull} (\algo) policy, see \Alg{alg:policy}. 
We consider a cutoff-based strategy for the following reasons: i)~the \DM should somehow define a value above which a \candidate might be accepted, value that needs to be consistent with the current \candidate sample (and not necessarily with the \refset) hence the need to explore before making any decision, ii)~in a finite-horizon settings with limited and constrained budget, the \DM should not rush into hiring since decisions are irrevocable, iii)~exploring the sample before making any decisions helps to estimate the quality of the \refset when we do not make the assumption that it is given to the \DM, and iv)~the intriguing behavior of the learning phase when the \refset has a given quality raised our curiosity.

Two other points concerning the cutoff-based \algo strategy. First, in practical situations where the quality of the \refset is good enough, that leads to an optimal cutoff value of $c^*= 0$, \ie it degenerates to a non cutoff-based strategy. The second point is that despite its name, the cutoff value $c$ is not the only parameter involved (the \emph{quality threshold}, or simply \emph{threshold}, $\tau_j$, is another one, see \Definition{def:threshold}). However, we found that it is the most critical parameter in driving the performance of the proposed strategy. The cutoff value being one of the key parameter of \algo algorithm, the policy is written $\policy(c)$ and therefore the regret becomes $\rsymb(c)$.

\subsection{Acceptance threshold}

Derived from the learning phase, the \algo policy dictates a set of \emph{threshold values} specific to each job position (\ie specific to each \referent that filled them) that \candidates need to exceed to be accepted. 
Since it depends on the available \referents, we first need to define the number of resignations by $r \in \{1,...b\}$, and therefore  $r := b - \Aobold{0}\scalar \oneS{b}$. The available \referents' scores are then denoted by $\mathbf{\symb{S}}^+ = (\symb{S}_{(i)})_{i\in I}$, where $I =  \{i : \Ao{i}{0} =1\}$, and is thus of size $b-r$.

In practice, during the selection phase, the acceptance threshold for each candidate is set to be the score of the $b$-th best up to the end of the learning phase. This set, called \emph{\updref}, is defined as $\Xrefbold{} = (\Xref{i}{})_{i \leq b}$ where each term belongs to the concatenation of both the \referents and the rejected candidates, \ie the $c$ first candidates, hence
$\Xref{i}{c} \in (\Sobold, \S{1},...,\S{c})$ \st $\Xref{1}{} > ... > \Xref{b}{c}$.
The threshold is a fixed value, which might not be optimal when every empty job positions have been filled. In fact, in the latter case, the threshold should be adapted to the scores of the available \referents, so that no position gets filled by a worse item.
Note that, during the learning phase candidates are rejected by default, hence the acceptance threshold is defined only during the selection phase.
Under these conditions, the acceptance threshold is defined as follows.

\begin{definition}{Step-specific acceptance threshold ($\tau_j$):} Score value to beat at step $j> c$ of the \WSSP when the \algo policy is applied with cutoff value $c$:
\begin{equation}
\tau_j(c) := 
\begin{cases}
{\Xref{b}{c}} & l < {\nres} + \sum_{\t=1}^{c} \Ind{\S{j} \ge \Xref{b}{c}};\\
 {\symb{S}^+_{(b-l)}}  & \text{otherwise},
\end{cases}
\end{equation}
where $l =  \sum_{i=c+1}^{j-1}A_i$. The second term in the condition is the number of \candidates from the learning phase that have been added in the \updref.
\label{def:threshold}
\end{definition} 

\begin{algorithm}[t]
\footnotesize
\caption{The proposed \algofull policy for \WSSP}
{\bf Input:} the number of $\budget$ jobs, the number of \candidates $n$, the number of resignations $\nres$, the \refset \scores from best to worst $\Sobold = (\So{1},\mydots, \So{b})$, the initial vector of \refset availability $\Aobold{0} = (\Ao{1}{0}, \mydots, \Ao{b}{0})$, and the cutoff value $c$.

{\bf Output:} the set of final job assignment $(\Aobold{n}, \Abold)$
\begin{algorithmic}[1]
\phase{Learning phase}\vspace{-0.7mm}%
\State $\Abold_{1,\mydots,c} \leftarrow 0$\algComment{reject by default all $c$ first \candidates}
\State $\mathbf{Y} \leftarrow \text{top\_of\_rank}(\budget, \,(\Sobold,\S{1},\mydots,\S{c}))$\algComment{$\budget$-best from $\Sobold$ and $(\S{1},\mydots,\S{c})$, in descending value order}
\State $n_{\text{rej}} \leftarrow \sum_{j=1}^c \Ind{\S{j} > \Xref{b}{c}}$\algComment{the number of \candidates among the $c$ first that beat...}
\Statex \algComment{...the threshold, \ie here, the last \rating of the \updref}
\State  $\symb{\mathbf{S}}^+ \leftarrow (\So{i})_{i\in I}$ where $I=\{i: \Ao{i}{\t}=1\}_{1\leq i \leq b}$     \algComment{initialize the selection with the available \refset}
\State $ l \leftarrow 0$ \hspace{5mm}\algComment{the number of jobs assigned so far in the selection}
\phase{Selection phase}\vspace{-0.7mm}%
\For{$j = c+1$ to $n$}
\If { $l < n_{\text{rej}}+\nres$ }\algComment{set the threshold that the $j$-th \candidate should beat (see \Definition{def:threshold})}
\State \ \ $\tau_j = Y_{(\budget)} $
\Else {\ $\tau_j = \symb{{S}}^+_{b-l}$}
\EndIf
\If {$l  < \budget$\,\,\,and\,\,\,($\S{j} > \tau_j$ or\,\,\,$ j-l = n-\nres+1$)}
\State $A_j \leftarrow 1$ 
\If { $l \geq \nres$ }
\State $\Ao{b-l}{j} \leftarrow 0$\algComment{remove job from \refset}
\EndIf
\State $l \leftarrow l+1$
\Else {\ \ $A_j \leftarrow 0$}
\EndIf
\EndFor
\end{algorithmic}
\label{alg:policy}
\end{algorithm}
\noindent Following the definition of the acceptance threshold, the decision variable is therefore given by:
\vspace{-1mm}
\begin{equation}
\A{j} = \Ind{j >c}\,\, \mathds{1}\left\{ \sum_{i=c+1}^{j-1}\A{i} < b\right \}\,\, \Ind{\S{j} \ge \tau_j},
\label{eq:accept}
\vspace{-1mm}
 \end{equation}
where the second indicator function ensures that no more than $b$ items can be selected. In the rest of the paper, the number of candidates accepted up to step $j$ (included) is denoted by $\aj = \sum_{i=1}^{j} \A{i}$. The \algo algorithm is fully described in \Alg{alg:policy}.
\begin{remark}Due to the finite horizon, the \DM might select candidates \emph{by necessity}, regardless their quality. This may occur in order to prevent having vacant positions in the output when the very end of the sequence is reached.
\end{remark}

\section{Optimal \algofull}\label{sec:theory}

We now propose an in-depth study of the properties of the cutoff strategies that takes advantage of the rank-based perspective used in the evaluation setup.

\subsection{Defining the quality}

A natural question that arises from the existence of the \refset concerns the `value' (or quality) of the \referents compared to the candidates next to come. How `good' is our initial set with respect to the arriving candidates? Besides, a notion of `good' should also be defined. We address the latter interrogation by introducing the `goodness' of $\Xobold$ for $\Xbold$, which we call \emph{quality of the \refset} and denote as $\quality$ (see \Definition{def:quality}). This parameter quantifies how the \refset ranks on average compared to the \candidates. Herein, we suppose that this parameter is provided in advance to the \DM. 
Other options to define the quality are possible but we found that the normalized average rank exhibits interesting properties.

\begin{definition}{True rank-based relative quality of \refset ($\quality$):} For a \WSSP, $\quality$ is the average normalized rank of the $\budget$ items of the \refset compared to the $n$ \candidates: 
\begin{equation}
\quality := 1- \frac{ \frac{1}{\budget} \Xobold\scalar\oneS{b} - \xmin}{\xmax-\xmin},
\end{equation}
where $\Xobold= (\Xo{1} =\rank(\So{1}),...,\Xo{b} = \rank(\So{b}))$ are the \referents absolute ranks, $\quality \in ]0,1[$, with $\quality \rightarrow 1$ as the \refset gets better skilled and $\quality=1/2$ corresponds to the medium quality \st $\frac{1}{\budget}\Xobold\scalar\oneS{b} = \frac{1}{2}(n+\budget+1)$.
\label{def:quality} 
\end{definition}

\subsection{Offline analysis}\label{offline_analysis}

\inlinetitle{Initialization}{.}
This analysis concerns the initialization of the process, \ie before the arrival of candidates, and is independent on the chosen strategy. The \DM has information about the average quality of the \referents, but we are particularly interested in the available ones, \ie  those with ranks $\Xavailbold = (\symb{X}_{(i)})_{i\in I}$, where $I =  \{i : \Ao{i}{0} =1\}$. These preselected \referents might end up, if competitive enough, in the final selection.

\begin{proposition} Let a given \WSSP starting with ${\nres} \le b$ resignations. The expectation of the rank of the $l$-th item from the available \refset $\Xavailbold$ is given by:
\begin{align} 
\Exp{\Xavail{l}} &= \frac{\gamma_0(\budget+1)l}{\budget(\budget-{\nres}+1)} \label{eq:gamma_0_res}, \\
\text{where }\, \ \gamma_0 := \Exp{\Xo{\budget}} &= (1-\quality)\frac{2\budget(n+\budget-1)}{\budget+1} +  \frac{2\budget}{\budget+1} \label{eq:gamma_0},
\end{align} 
\st $\gamma_0$ is the expectation of the $\budget$-th item from the \refset $\Xobold$, and a function of the relative quality $\quality$ of the \refset.
\label{prop:gamma_0_res}
\end{proposition}

\inlinetitle{Offline selection}{.} It is desirable for any online algorithm to perform as close as possible to the optimal \emph{offline case} where the \DM knows the $\budget$-best items and can directly select them. Hence, we want our strategy to converge towards the {offline case} and have $\rsymb$ as small as possible. 
The offline output $\coff \in \real_+$ is given by \Definition{def:wssp} as:
\begin{equation}
\coff := \underset{(\symb{\mathbf{\B}}_n, \,\mathbf{\B}) \in \policy_{\text{off}}{_{,\bckrnd}}}{\min} \left(\Xobold\scalar \symb{\mathbf{\B}}_n + \Xbold\scalar \mathbf{\B} \right),
\end{equation}
where $\policy_{\text{off}}{_{,\bckrnd}}= \left\{ (\symb{\mathbf{\B}}_n, \,\mathbf{\B}) \in \{0,1\}^{n+b} :\,  (\symb{\mathbf{\B}}_n,\mathbf{\B})\scalar \oneS{n+b} = b\right\}$.
\begin{proposition} In the \WSSP context, the expected minimal \regret an offline algorithm can achieve, by selecting the $\budget$-best out of the $n+\budget-\nres$ \candidates and available \referents, is:
\begin{equation}
\Exp{\coff} = \frac{b(\budget+1)}{2}+\frac{\nres b^2 (\gamma_0 + \nres)}{2\gamma_0^2},
\label{eq:r_opt}
\end{equation}
where $\gamma_0$ is given in \Proposition{prop:gamma_0_res}.
\label{prop:r_opt}
\end{proposition}
The first term of \Eq{eq:r_opt} accounts for the standard average offline \regret, \ie the sum of the $b$-best ranks, while the second term represents the increase due to potentially unavailable items from the $b$-best. 
 
\subsection{Optimal cutoff and \WSSP main parameters for $q=1/2$}\label{sec:opt_ccm}

Let us first consider that, on average, \referents have a medium quality \ie $q=1/2$. Indeed, the analytical computation of the main variables of the problem is more challenging when $\quality \neq 1/2$, therefore we provide what we call a \emph{translation method} to `translate' any setting of arbitrary $\quality$ to the situation where $\quality=1/2$ for which we have analytical results (see \Sec{sec:translation_method}).
\begin{lemma} Let a \WSSP with $n$ \candidates, and a \refset of size $\budget$. Using \Eq{eq:accept}, a candidate is accepted if his rank beats the rank-based threshold, $\gamma_j = \rank(\tau_j)$, and less than $b$ candidates have been accepted. The probability for the number of accepted candidates at step $j$ to be smaller than $\budget$ is given by:
\begin{align} \begin{split}
g_j(\budget) := \Prob(\ajm < \budget) = 
\begin{cases}
  1,& \budget > j-c-1; \\
  e^{-\lambda_{j-1}} \sum_{i=0}^{b-1}\frac{  \lambda_{j-1}^i}{i!} + o(\smallo) & \budget \leq j-c-1,
\end{cases}
\end{split} \end{align} 
where $\lambda_{j-1} = \sum_{i=c+1}^{j-1}\pind $ and $\smallo = \sum_{i=c+1}^{j-1} (\pind )^2$.
\label{lem:g}
\end{lemma}
\begin{theorem} Applying the \algo algorithm with parameter $c$ as cutoff value, given that $\nres$ \referents resigned, and using \Lemma{lem:g}, the \WSSP exhibits the following features:
\begin{itemize}
\item Expected rank-based acceptance threshold for candidate $j$ is given by $\gamma_j := \Exp{\rank(\tau_{j})}$ \st:
\begin{align}
\gamma_j &= \gamma g_j(\Delta) + \frac{\gamma_0(b+1)}{b(b-r+1)}\left(\budget-\sum_{i=1}^{j-1}\pind g_i(\budget)\right)(1-g_j(\Delta) ),\label{eq:gamma_j_q1/2}
\end{align}
where $\gamma := \Exp{\rank(\Xref{b}{j})} = \frac{\budget(\budget+n)}{\budget+c}$, $\Delta = {\nres} + c\frac{\gamma-1}{n+b}$ and $\gamma_0$ is given in \Proposition{prop:gamma_0_res}. 
\label{prop:gamma_j_q1/2}
\item Expected number of new hires at the end of the selection $\An \le b$:
\begin{equation}
\Exp{\An} = 
 \sum_{j=1}^n \pjnd g_j(b)
\end{equation}
\item Expected \regret function to minimize, \ie expected average rank of the selected items:
\begin{align}
\begin{split}
\Exp{\rsymb(c)} &= \frac{1}{ (n+b)} \sum_{j = c+1}^{n} g_j(\budget) \frac{\gamma_j(\gamma_j-1)}{2} + \frac{\gamma_0(\budget+1)}{2\budget(\budget-{\nres}+1)} (\budget- \Exp{\an})(\budget+1 - \Exp{\an}) - \Exp{\coff},
\end{split}
\label{eq:expectation}
\end{align}
where $\Exp{\coff}$ is the expected minimal offline \loss defined in \Proposition{prop:r_opt}.
\end{itemize}
\label{th:expected_cost}
\end{theorem}
\Eq{eq:expectation} holds a good approximation of the expected \regret of \WSSP when $\quality=1/2$.
Recall that we want to find the optimal cutoff value 
$c^* =  \underset{c}{\text{argmin }} \Exp{\rsymb(c) }$ 
which is equivalent to finding $c^*$ \st $\frac{\partial}{\partial c} \Exp{\rsymb(c)} |_{c=c^*} = 0$. Unfortunately, this equation is analytically intractable unless approximations or restrictive assumptions are made, however we can easily spot $c^*$ numerically by tracking the lowest point of the curve $\Exp{\rsymb(c)}$ using \Eq{eq:expectation}, $\forall b, \forall n$, and store the results in $c^*(\budget,r,n)$.
\begin{remark}
 Note that in practice $\an$ is actually equal to $\max(\an, {\nres})$ to avoid empty positions at the end of the selection. An approximation of $\Exp{\max(\an, {\nres})}$ can be found in the Appendix, as well as an empirical verification.
\end{remark}
\inlinetitle{Example}{.}
Imagine a \WSSP instance where $n=100$ candidates are going to be sequentially interviewed. The \DM handles $b=5$ job positions, each of them already filled by available \referents (\ie $\nres=0$) of a given quality $q=0.75$ \wrt to the candidates next to come. Using \Theorem{th:expected_cost}, the length of the learning phase is $c^*=\lfloor 38.36\rfloor = 38$, the expected average rank of the selected items is $\Exp{\rsymb(c^*)}/5 = 2.60$, and the expected number of accepted candidates is $\Exp{\An} = 0.997$. The latter is low since the initial quality is quite good, hence the \DM expect to fire only his worse \referent (available \referent). Now, with the same setup of \WSSP parameters except with full resignations, \ie $r=b$, we get $c^*=\lfloor 28.27\rfloor = 28$, $\Exp{\rsymb(c^*)}/5 = 3.40$ and $\Exp{\An} = 5$, which is coherent with the fact that all positions are initially empty. The length of the learning phase is reduced compared that of the previous example, implying a less competitive acceptance threshold. Justifiably, the \DM is less demanding on the quality of the accepted items, to avoid the risk of having to select last incoming candidates by default, called a \emph{failure} (see \Sec{sec:adjusted_algorithm}).
\begin{figure}[t]\centering
\hspace{-5mm}
\subfigure{
\clipbox{2 6.3pt 0 0}{
\includegraphics[width = 0.245\linewidth, viewport=25 0 565 600, clip]{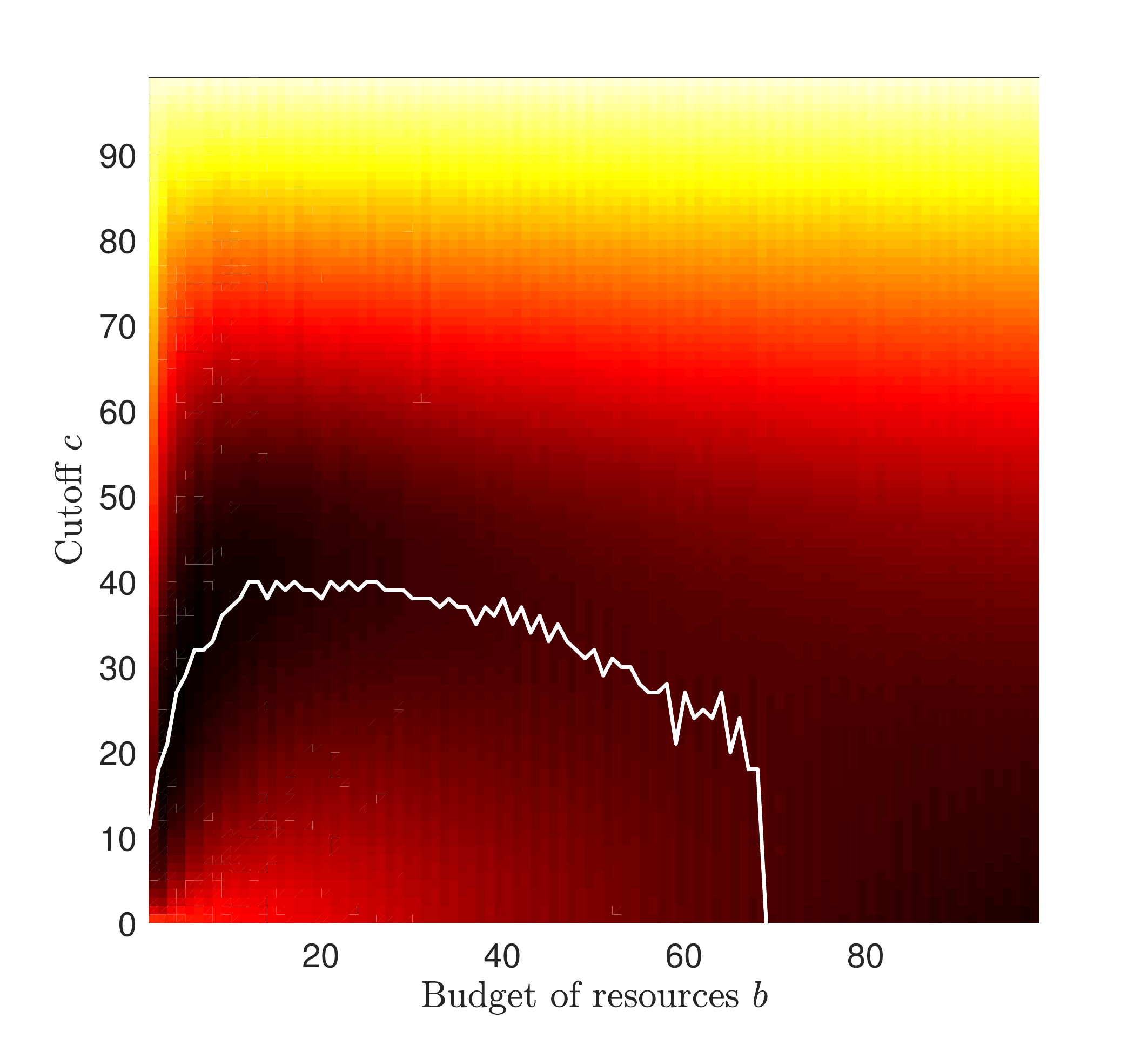}} }\hspace{-0.4mm}%
\subfigure{
\clipbox{8pt 6.3pt 0 0}{
\includegraphics[width = 0.245\linewidth, viewport=25 0 565 600, clip]{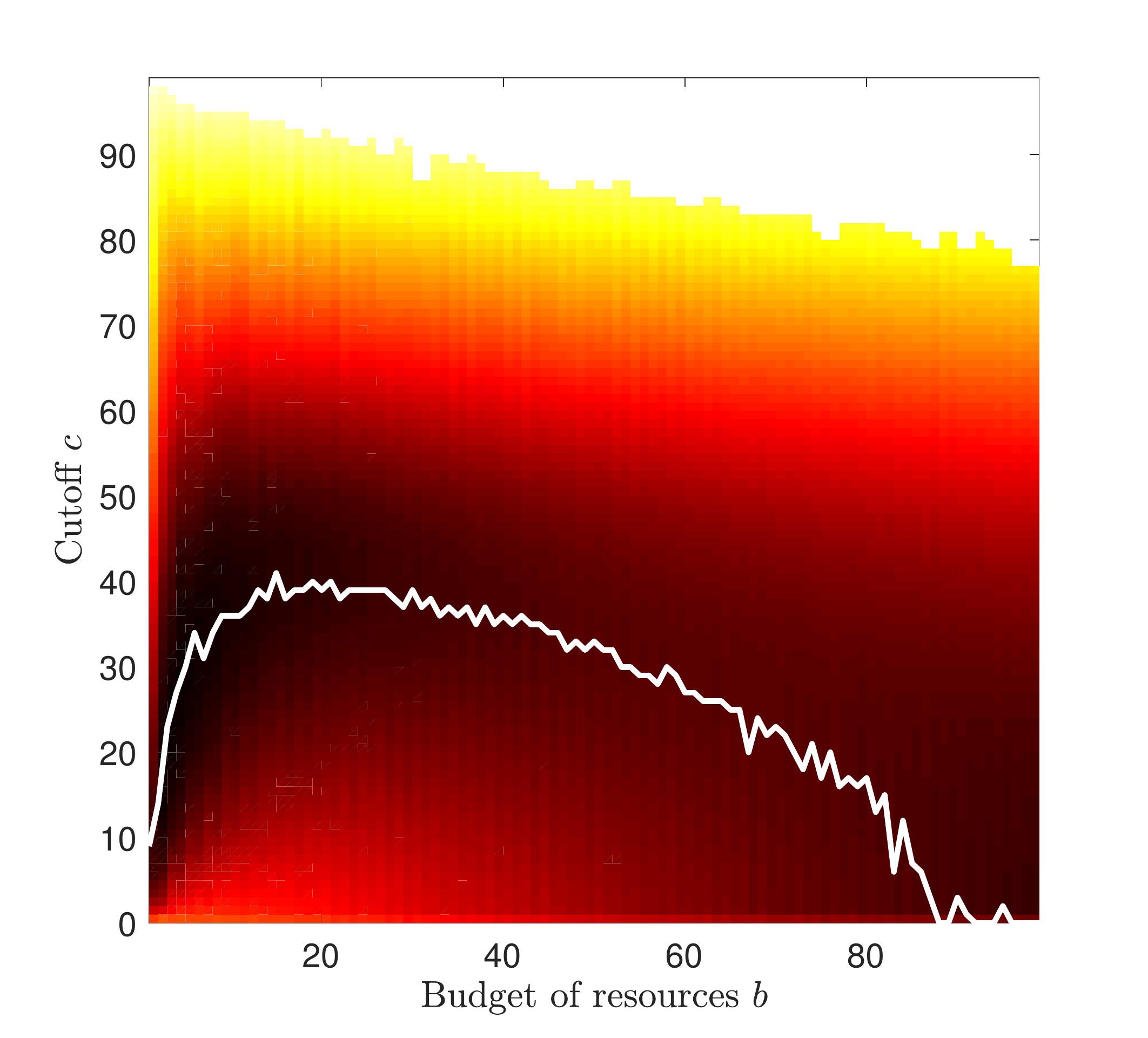}} }\hspace{-0.5mm}%
\subfigure{
\clipbox{8pt 6.3pt 2pt 0}{
\includegraphics[width = 0.245\linewidth, viewport=25 0 565 600, clip]{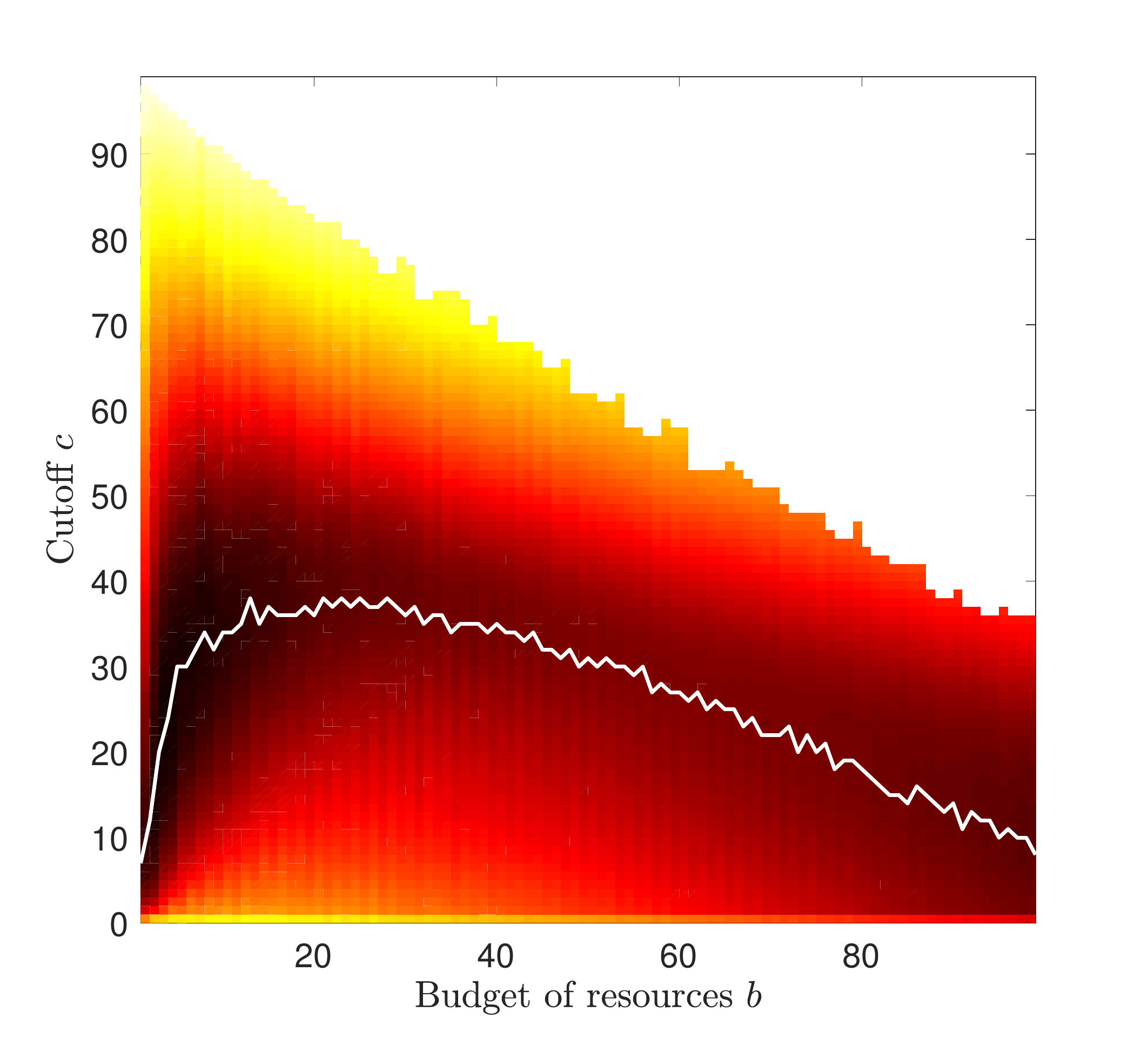}} }\hspace{-0.4mm}%
\subfigure{
\clipbox{8pt 6.3pt 0 0}{
\includegraphics[width = 0.262\linewidth, viewport=25 0 605 600, clip]{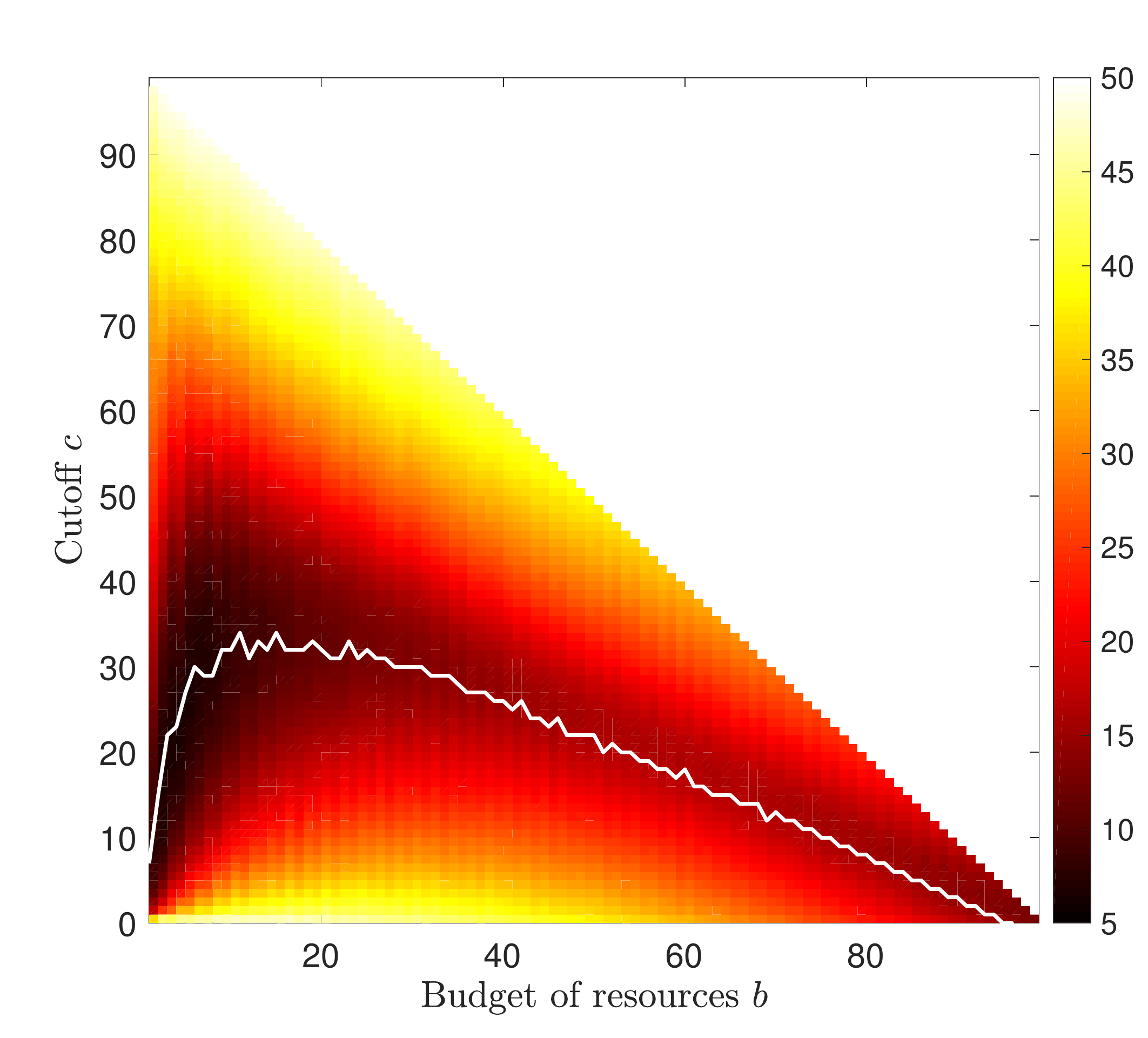}}}\\ 

\vspace{-7mm}%
\hspace{-5mm}%
\setcounter{subfigure}{0}
\subfigure[\small $\nres = 0$]{
\clipbox{2 3pt 0 0}{
\includegraphics[width = 0.245\linewidth, viewport=25 0 565 600, clip]{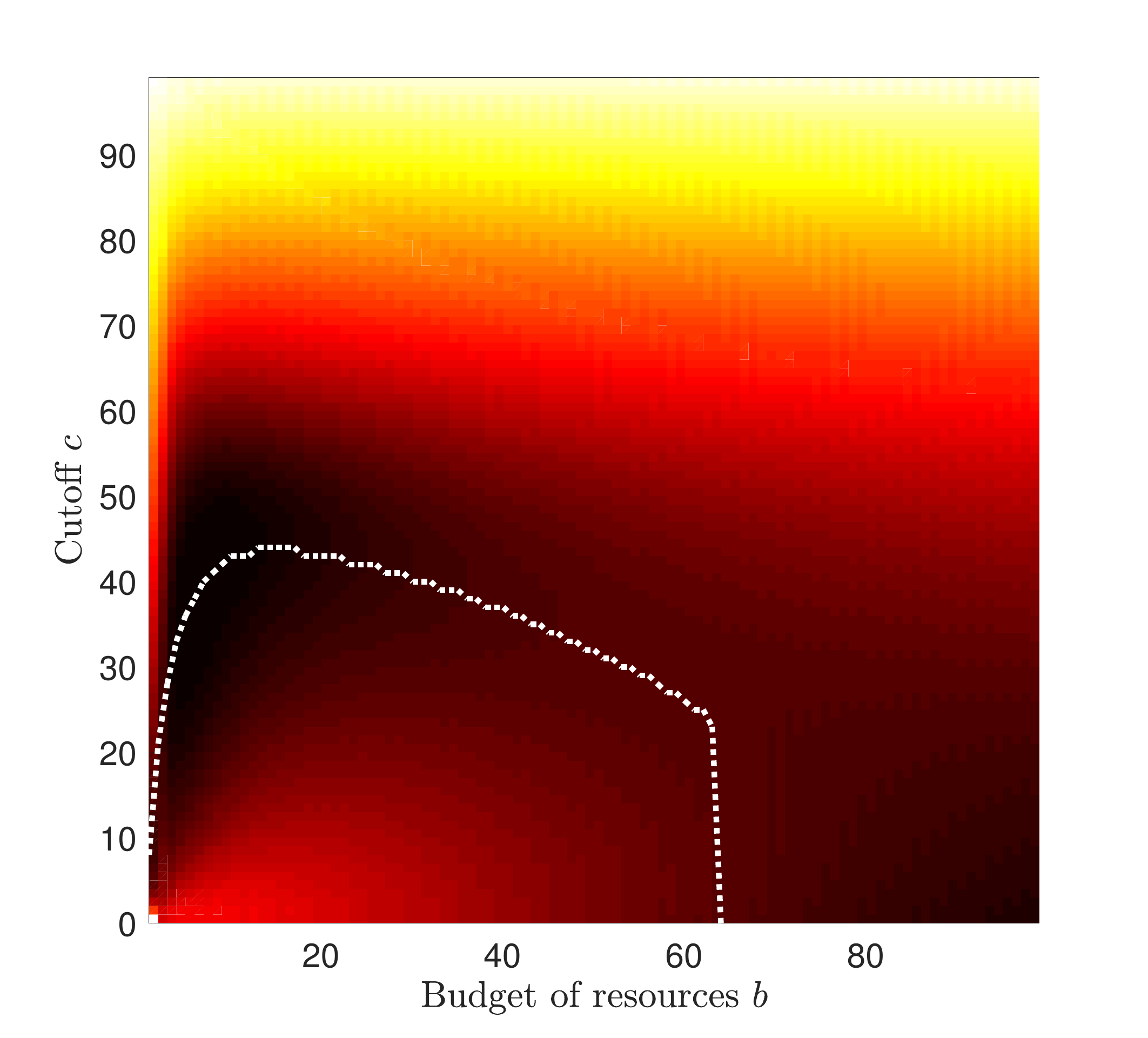}} }\hspace{-0.4mm}%
\subfigure[\small $\nres = 0.1b$]{
\clipbox{8pt 3pt 4pt 0}{
\includegraphics[width = 0.245\linewidth, viewport=25 0 565 600, clip]{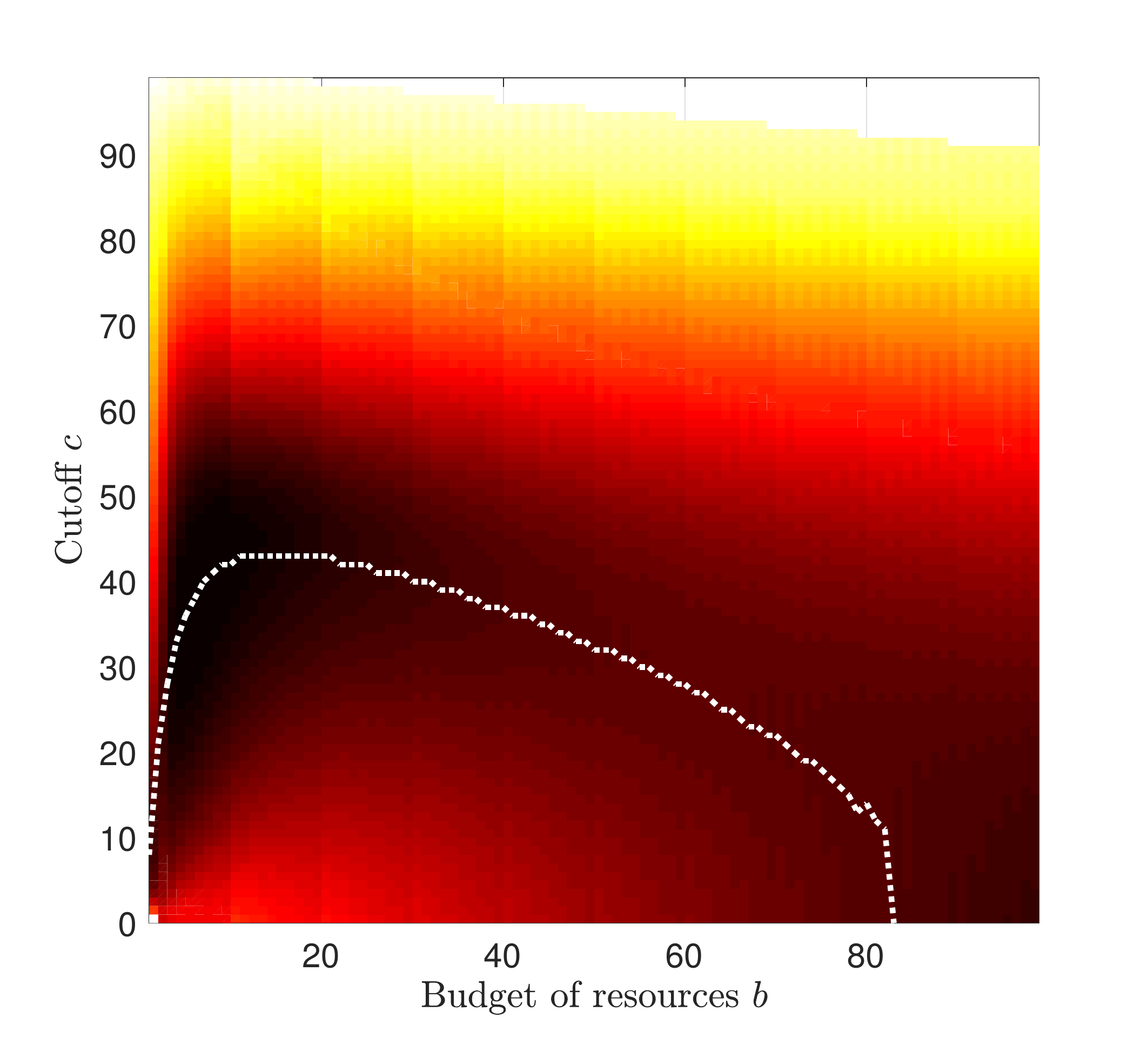}}}
\hspace{-0.4mm}%
\subfigure[\small $\nres = 0.5b$]{
\clipbox{9pt 3pt 4pt 0}{
\includegraphics[width = 0.245\linewidth, viewport=25 0 565 600, clip]{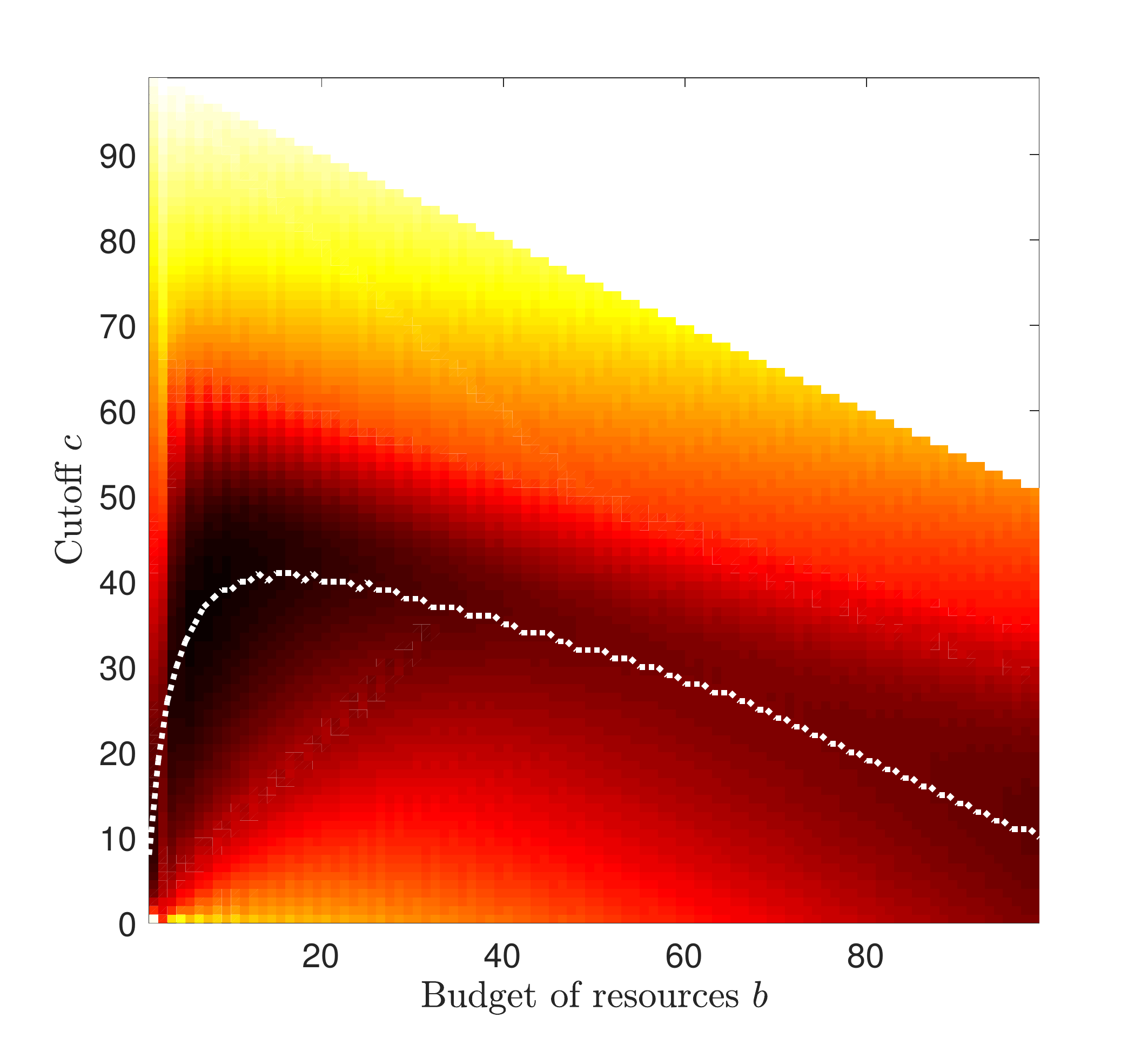}}}  
\subfigure[\small $\nres = b$]{
\clipbox{9pt 3pt 0pt 0}{
\includegraphics[width = 0.262\linewidth, viewport=25 0 605 600, clip]{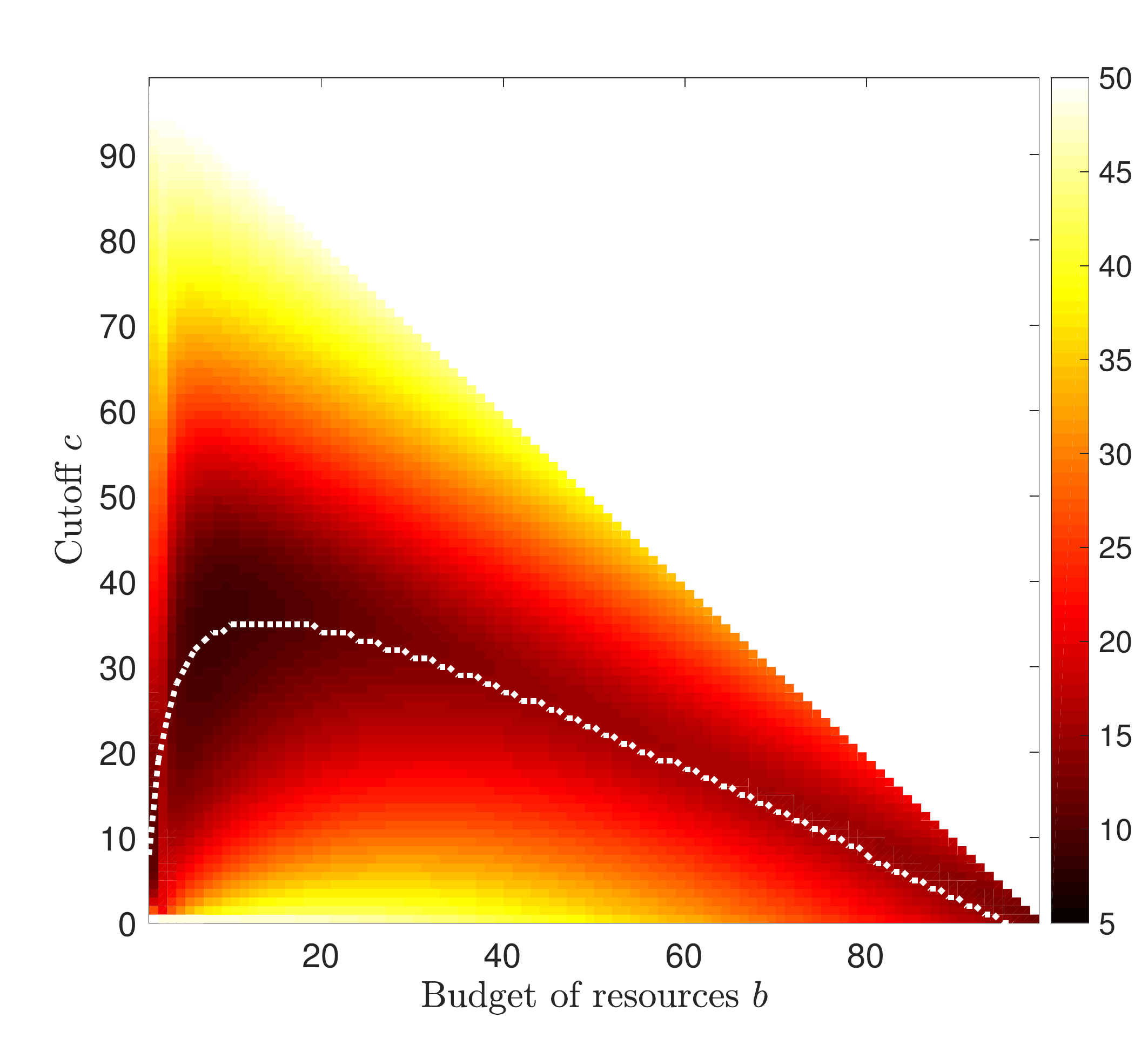}} }\hspace{-0.5mm}%
\caption{\captionSize Comparative heatmaps for the average empirical \regret (top line) and the expected \regret (bottom line) derived from \Theorem{th:expected_cost} over different resignation numbers $\nres = \{0, 0.1b, 0.5b, b\}$, and both for \refset quality $\quality=1/2$. In each case, the heatmap of the \regret is presented over the parametrization of the cutoff value $c$ and the number of jobs $\budget$ (budget).}
\label{fig:law_large_number}
\end{figure}

\inlinetitle{Simulations}{.}

In order to guarantee the accuracy of our analytical approximation $\Exp{\rsymb(c)}$ in \Eq{eq:expectation}, we simulate 
each \WSSP scenario 1000 times : for a fixed number of \candidates $n=100$ and a fixed \refset quality $\quality = 1/2$.  
The top row of \Fig{fig:law_large_number} displays a heatmap of the average empirical \regret (simulated) \wrt the number of jobs $\budget$ (x-axis) and the value of the cutoff $c$ (y-axis). The white plain line in each heatmap follows the path of the lowest simulated value of the heatmap, referred to as $c^*_\text{sim}(\budget) = c^*_\text{sim}$. These plots should be put in comparison with those in the bottom row which show the heatmaps of the expected \regret according to our analysis. The white dashed line follows again the path of the lowest heatmap value, which we denote as $c^*(\budget) = c^*$.
From \Fig{fig:law_large_number}, it becomes clear that the law of large number complies with the lemmas and propositions of \Sec{sec:opt_ccm} which are consistent in these experiments.

\subsection{Optimal cutoff for arbitrary $q$}\label{sec:translation_method}

\subsubsection{The translation method}

In \Sec{sec:opt_ccm}, we derived an analytical expression for $\Exp{\rsymb(c)}$ given a relative quality of the \refset $\quality=1/2$. However, when $\quality \neq 1/2$, the analytical computation of the \WSSP's main variables is highly complex. We introduce a rather simple trick to efficiently overcome this difficulty. More specifically, we provide a way translate any setting of arbitrary $\quality$ to a \emph{$\gamma_0$-similar setting} where the quality of the \refset is set to be $\quality=1/2$ and for which we can use the results presented in \Sec{sec:opt_ccm}. We introduce a notion of \emph{similarity} between two different settings' \refset $\gamma_0$ (see \Definition{def:translation}) and come up with what we call as the \emph{translation method} described below (see \Proposition{prop:translation}).

\begin{figure}[t]
\centering
\hspace{-3mm}
\subfigure[$\nres=0$]{\includegraphics[width = 0.36\linewidth]{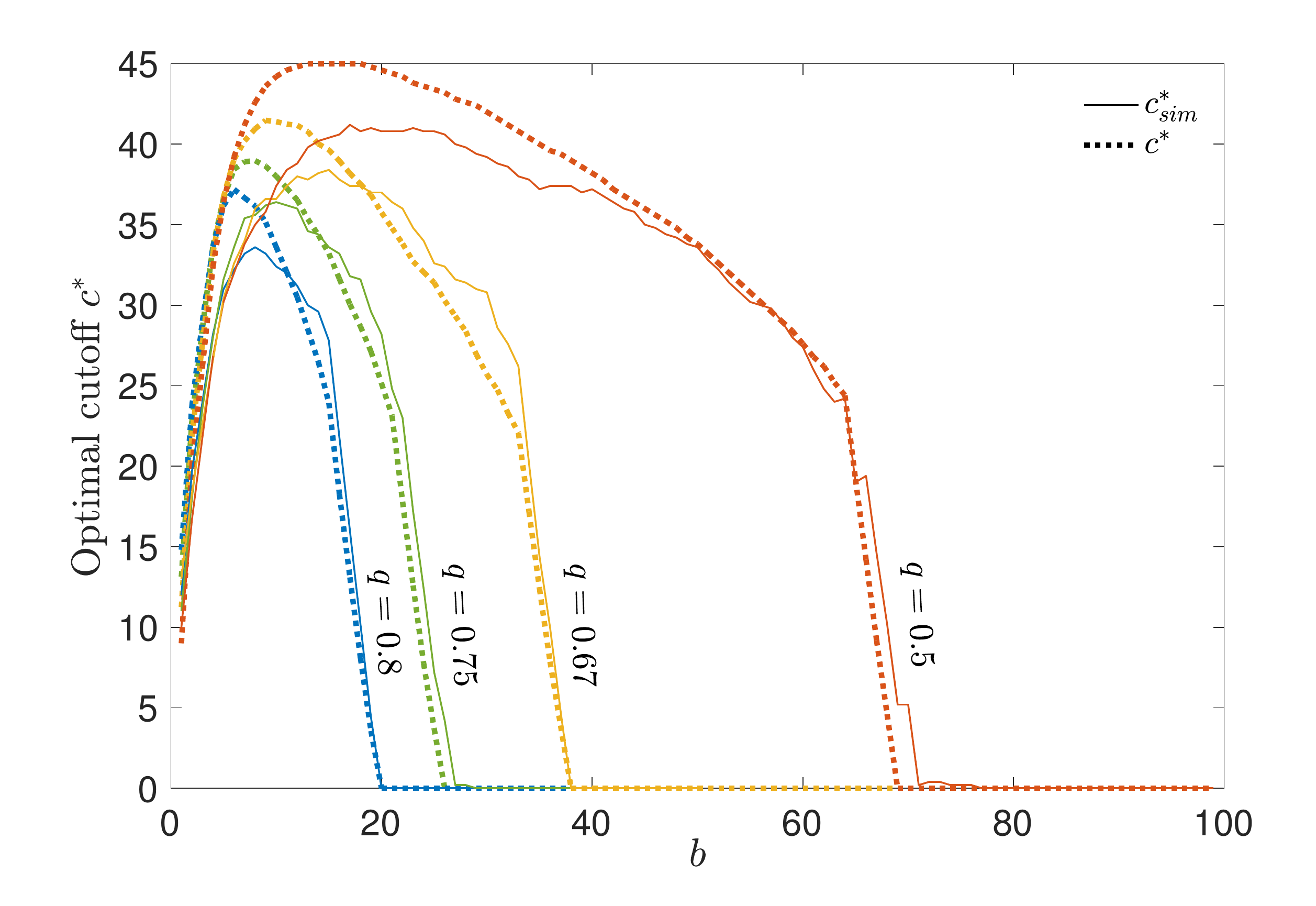}}
\hspace{-4mm}
\subfigure[$\nres=0.5b$]{
\clipbox{17pt 0pt 3pt 0pt}{
\includegraphics[width = 0.353\linewidth]{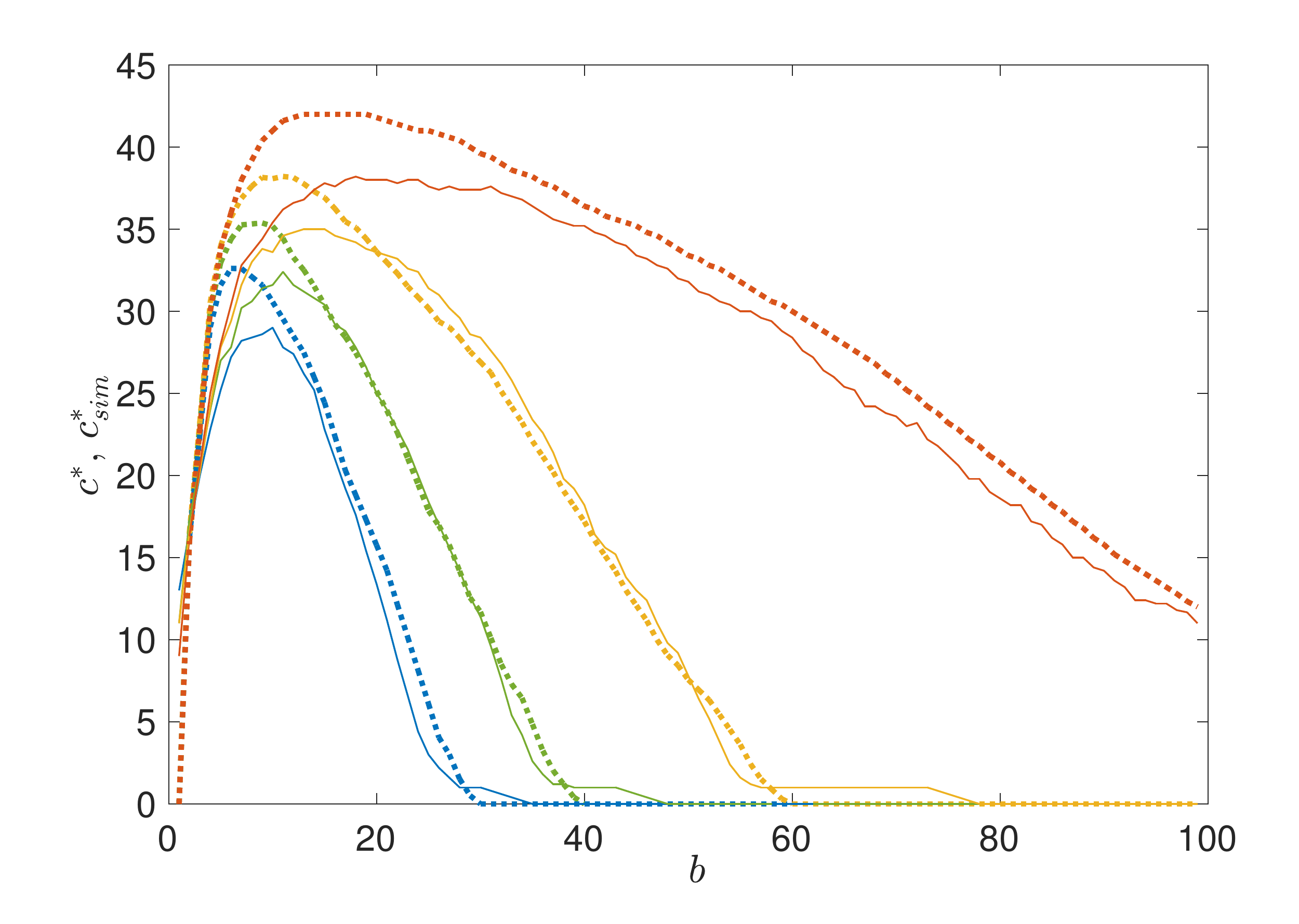}}}
\hspace{-3mm}
\subfigure[$\nres=b$]{
\clipbox{17pt 0pt 0pt 0pt}{
\includegraphics[width = 0.353\linewidth]{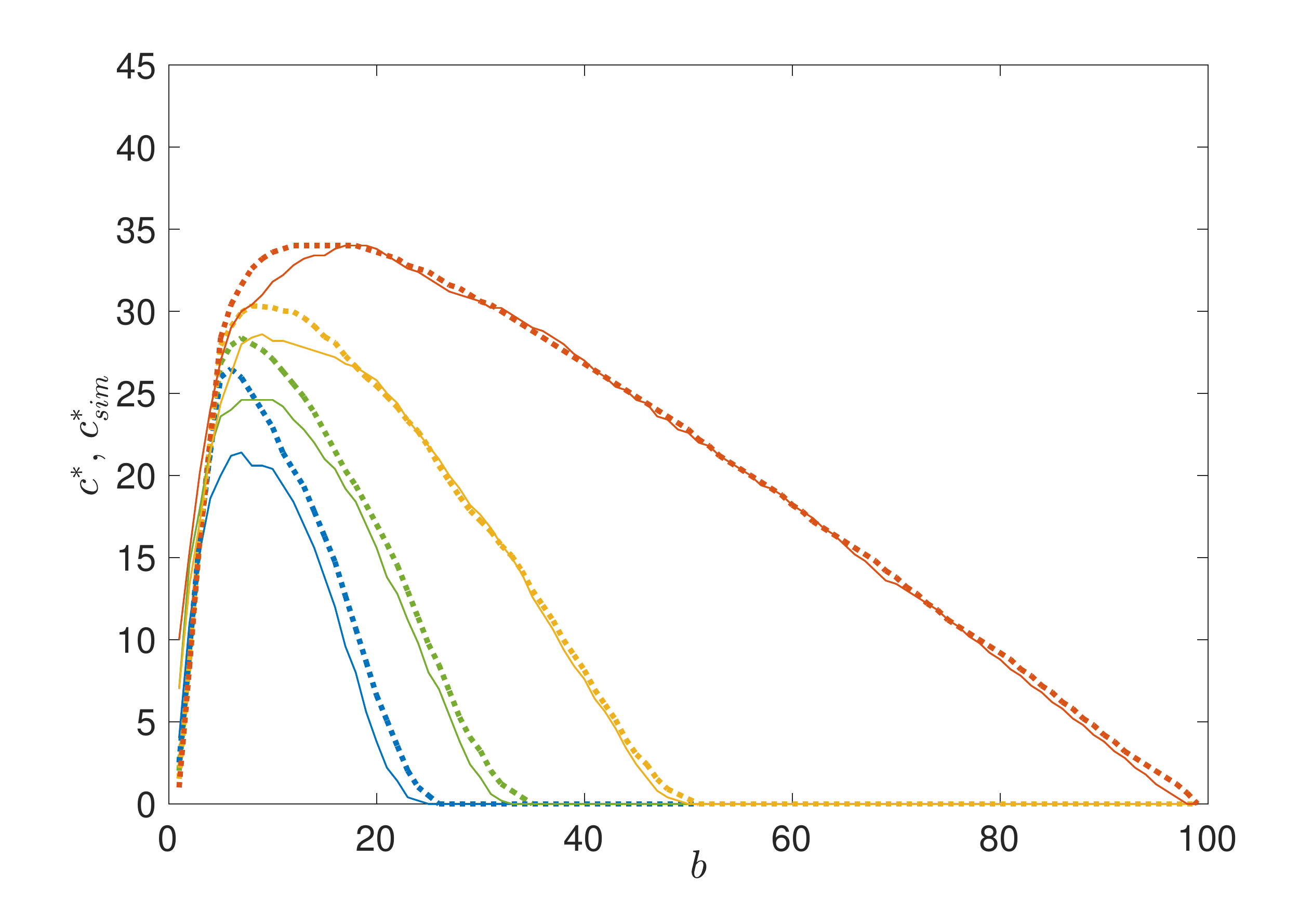}}}
\caption{\captionSize The optimal cutoff \wrt the number of jobs $\budget$ (x-axis), according to the simulations in plain lines and to our analytical approximation (see \Eq{eq:cost}) in dashed lines for different values of the relative quality of the \refset $\quality = \{\frac{1}{2}, \frac{2}{3}, \frac{3}{4}, \frac{4}{5}\}$, for $n=100$ \candidates.}
\label{fig:translation_method}
\end{figure}

\begin{definition}{$\gamma_0$-similarity}: Suppose each \WSSP instance, denoted by \WSSP\!\!$_x$, starts with $\budget_x$ jobs positions filled with the available \referents $\Xavailbold_{x}$, and thereafter interviews $n_x$ \candidates using \algo (see \Alg{alg:policy}) with the optimal cutoff value $c^*_x$. Then, the settings of two instances, \WSSP\!\!$_x$ and \WSSP\!\!$_y$, are said to be $\gamma_0$-similar if their \refsets (even those unavailable) have the same ranks \wrt the rest of the sample, regardless $n_x$ and $n_y$:
\begin{equation}
\text{\WSSP}_x \ \equiv_{(\gamma_0)} \text{\WSSP}_y \hspace{3mm}\text{if} \hspace{3mm}
\begin{cases} 
\budget_x = \budget_y, & \\
{\nres}_{x} = {\nres}_{y}, &\\
\Xobold_{x} = \Xobold_{y}  \Rightarrow \gamma_{0;x} = \gamma_{0;y}.
\end{cases}
\end{equation}
\label{def:translation}
\end{definition}
\begin{proposition} Translation method: If $\text{\WSSP}_x \ \equiv_{(\gamma_0)} \text{\WSSP}_y$, then: 
\begin{equation}
c^*_y = c^*_x\,\frac{n_y+\budget}{n_x+\budget} \, \ \ \text{and}\, \ \ n_x = (n_y+\budget-1)\frac{1-\quality_y}{1-\quality_x} - \budget+1.
\end{equation}
\label{prop:translation}
\end{proposition}
\vspace{-6mm}
\Alg{alg:translation} describes the overall translation algorithm. The plots in \Fig{fig:translation_method} indicate the large agreement between the optimal cutoffs computed by our analytical translation method and the cutoff empirically computed through simulations.

\begin{algorithm}[t]
\footnotesize
\caption{Translation method between two WSSP settings}
{\bf Input:} the \WSSP setting of interest, $\text{\WSSP}_t$ (subscript $t$ for `\emph{target}' and, below, $s$ for `source'), and its main parameters: $n_t$ \candidates, $\budget_t$ jobs, a \refset of a relative rank-based quality $\quality_t$ from which ${\nres}_{t}$ resigned, and the vector $c^*(\budget_t,n)$ with the optimal cutoffs for any sequence length $n$, for $\quality=1/2$, as described in \Theorem{th:expected_cost}.

{\bf Output:} the optimal cutoff $c^*_t$ for the $\text{\WSSP}_t$ setting.
\begin{algorithmic}[1]
\phase{Find a $\gamma_0$-similar setting to $\text{\WSSP}_t$, let that be $\text{\WSSP}_s$}\vspace{-0.7mm}%
\State Require $\budget_s = \budget_t = \budget$, as in \Definition{def:translation}
\State Impose $\quality_s = 1/2$
\State Compute $n_s = \lfloor (n_t+\budget-1) \frac{1-\quality_t}{1-\quality_s}-\budget+1 \rfloor$, as suggested by \Proposition{prop:translation}
\phase{Translate the setting $\text{\WSSP}_s$ to $\text{\WSSP}_t$}\vspace{-0.7mm}%
\State Find the best cutoff value $c^*_s = c^*(\budget,n_s)$ from the input vector
\State Compute $c^*_t =  \lfloor c^*_s\,\frac{n_t+\budget}{n_s+\budget} \rfloor$, according to \Proposition{prop:translation}
\end{algorithmic}
\label{alg:translation}
\end{algorithm}
\inlinetitle{Examples}{.} 
Let us illustrate the translation method with one example. Imagine the \DM deals with $\text{\WSSP}_t$, where no \referent resigned, with $n_t=100$, $\budget_t=15$, $\quality_t=0.8$, and she is interested in knowing $c^*_t$. One possible $\gamma_0$-similar setting, $\text{\WSSP}_s$, has the following features $\quality_s = 1/2$ and $\budget_s=\budget_t=b=15$. Using \Proposition{prop:translation} we get $n_s = \lfloor (n_t+\budget-1) \frac{1-\quality_t}{1-\quality_s}-\budget+1 \rfloor = \lfloor 
\frac{114 \cdot 0.2}{0.5} - 14 \rfloor = 31$; then using \Theorem{th:expected_cost} we compute $c^*_s$ numerically for $n_s=31$ (which is feasible as long as $\quality_s=1/2$) and get $c^*_s=9$. Finally we obtain $c^*_t= \lfloor c^*_s\,\frac{n_t+\budget}{n_s+\budget} \rfloor = 22$; the \DM rejects the first $\frac{c^*_t}{n_t}$ \candidates, that is $22\%$ of the total sample, before starting to select.

\inlinetitle{Simulations}{.}
For a fixed quality $\quality$, it is worth pointing out that $c^*(\budget)$ is not a monotonic function but rather has two distinct regimes indicated by the $\text{sign}(\frac{\partial c^*}{\partial \budget})$. This can be better explained as the following trade-off. %
Suppose fixed $n$ and $\nres$ (see \Fig{fig:translation_method}) and that we start with $b=1$: increasing $\budget$ would mean more jobs to assign, hence, the \DM should very quickly (\wrt budget increase) increase the length of the rejection phase to make sure that she learns sufficiently before taking the many decisions (regime $\frac{\partial c^*}{\partial \budget} \geq 0$). From a point and further, though, increasing $\budget$ would also mean a)~to have a less competitive threshold (which depends on the quality of the worst current \referents), b)~that the whole process becomes less selective as less and less \candidates need to be rejected, c)~to have a higher expected number of resignations (if $\nres > 0$), which makes the exploration for the \DM less safe. Hence, the \DM should start shortening her learning phase (regime $\frac{\partial c^*}{\partial \budget} < 0$). The optimal cutoff values get lower as the number of resignations $\nres$ increases (see the curves across the plots of \Fig{fig:translation_method}), as well as with the decrease of \refset quality $\quality$ (see the compared curves in each plot of \Fig{fig:translation_method}). 

\section{Adjusted policy: \algoNewfull}\label{sec:adjusted_algorithm}

In real-life scenarios the proportion of \referents that resign compared to those who stay is often relatively small; therefore, in the presented recruitment context, the more relevant results of this work concern situations where the number of resignations is small (\eg 2$\nres \leq b$). 
However when the latter is quite high (\ie most job positions are empty), the \DM might have to accept last arriving \candidate(s) in order to fill vacant positions, this event is called a \emph{failure} (described in \Definition{def:failure}) and is similar to hiring random \candidate(s) which ends up increasing the \regret.

\begin{definition}{Failure and failure rate ($f_j$):} A failure at step $j$ is the event of accepting a last incoming \candidate by default (to fill empty job positions) whose \score did not beat its associated threshold $\tau_j$, \ie
$f_j = \Ind{j - \ajm = n - {\nres} + 1}\Ind{\S{j} < \tau_j}$. The failure rate $\rho_f$ is defined as the sum of the number of failures divided by the number of tests.  
\label{def:failure}
\end{definition}

\begin{figure}[t]
\centering
\subfigure[$\pres = 0.5$ and $\rsymb(c^*)=0.8$ ]{\includegraphics[width = 0.34\linewidth]{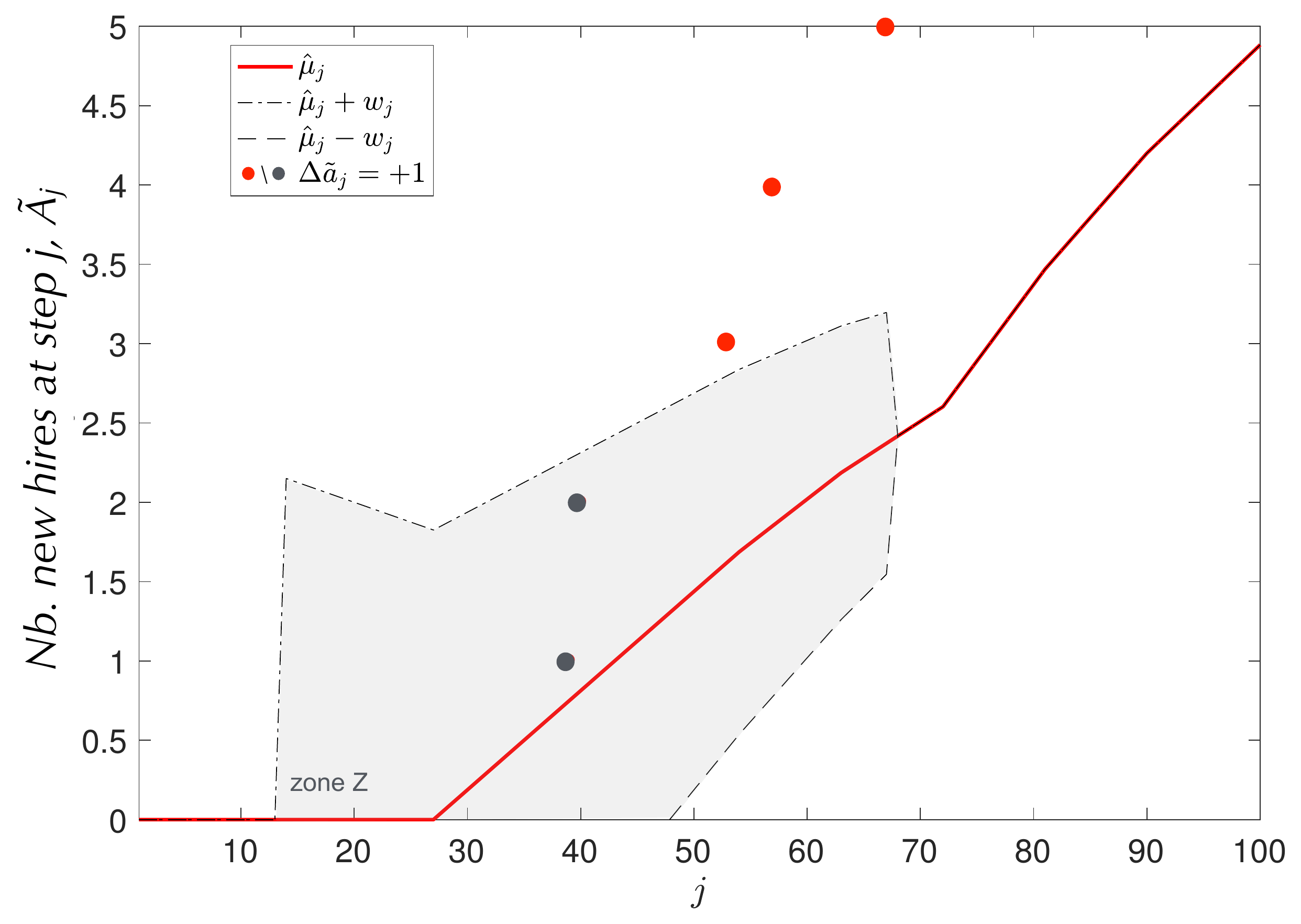}}%
\subfigure[$\pres = 0.8$ and $\rsymb(c^*)=2.4$]{
\clipbox{11pt 0pt 0pt 0pt}{
\includegraphics[width = 0.34\linewidth]{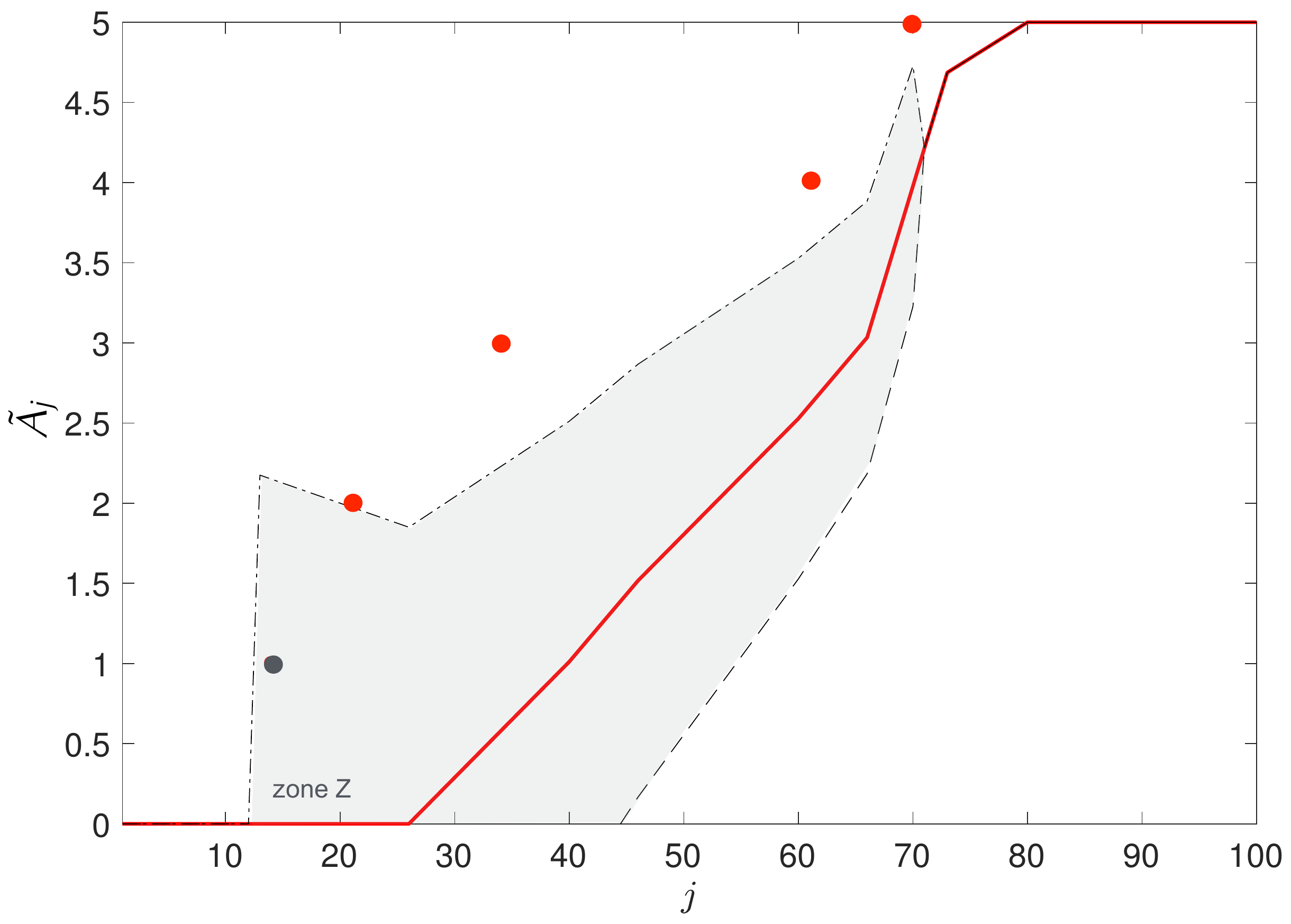}}}%
\subfigure[$\pres = 1$ and $\rsymb(c^*)=1$]{
\clipbox{11pt 0pt 0pt 0pt}{
\includegraphics[width = 0.34\linewidth]{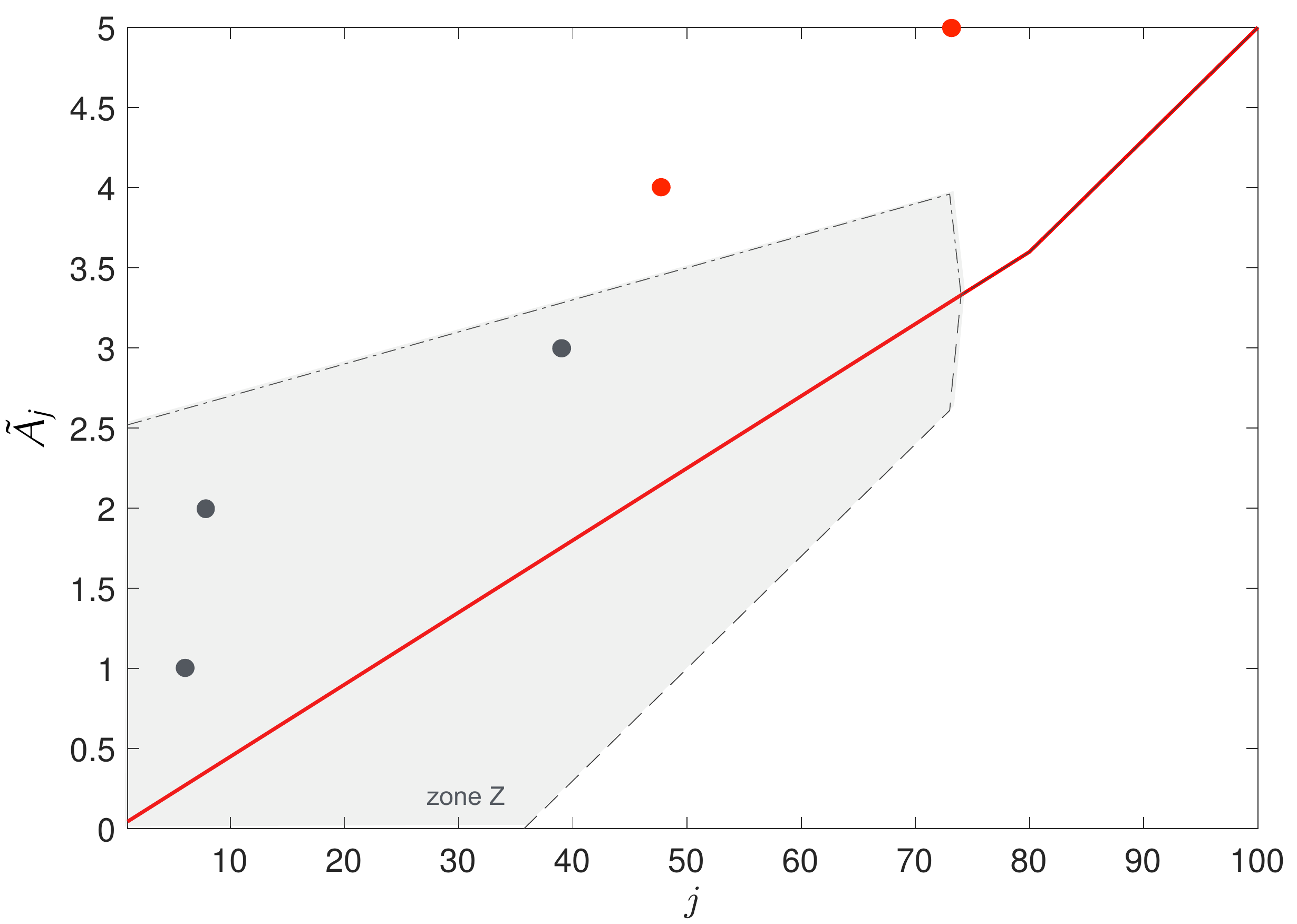}}}
\caption{\captionSize Expectation of the number of accepted items at step $j$ \wrt the step $j$ for $n=100$ \candidates given that no \candidate was accepted by default (red curve). Number of jobs $b=5$. We simulate a single \WSSP, gray shade and red points indicate the index $j$ of an accepted \candidate. The zone (see \Definition{def:zone}) is gray-shaded with $w_j=\frac{\budget}{2}\left(1-\frac{j}{n}\right)$. Gray points stand for \candidates that were accepted using a constant threshold (they lay inside the zone $Z$), and red points using the \algoNew variation.}
\label{fig:zone}
\end{figure}

\noindent Simulations show that in some settings the failure rate is indeed significant, for instance it reaches $\rho_f = 0.58$ for $b=20$, ${\nres}=b$, and $q=0.81$. This phenomenon appears due to the high quality of the \updref, \ie the threshold becomes too competitive and hence difficult to beat for most \candidates. Our idea to mitigate this effect is to estimate the expected number of accepted \candidates at step $j$, denoted by $\hat{\mu}_j(c, {\nres})$, given that the total number of accepted \candidates (\ie at the end of the selection) is greater or equal to the number of resignations ${\nres}$; formally that is: $\hat{\mu}_j(c, {\nres}):= \Exp{\aj \mid \an \geq {\nres}},\, \ \forall j$, \ie there is no failure (see \Proposition{prop:mu_hat}).%
\begin{proposition} The expectation of the number of \candidates accepted at step $j$ given that there is no failure is given by: 
\begin{equation}
\hat{\mu}_{j} (c, {\nres}):= \Exp{\aj \mid \an \geq \nres} =  \lambda_j g_{j+1}(b-1) +  \frac{\budget(1-g_{j+1}(\budget))}{1-g_{n+1}({\nres})},
\end{equation}
where $\lambda_j = \sum_{i=c+1}^j \pind$ and $g_j(x) :=\Prob(\ajm < x)$ (see \Lemma{lem:g}).
\label{prop:mu_hat}
\end{proposition}
The proof is detailed in the Appendix. We use \Proposition{prop:mu_hat} to compute $\hat{\mu}_{j} (c^*, {\nres}),\, \ \forall j \in \INTerval{1}{n}$, and compare it to the current number of accepted \candidates at step $j$ (included), denoted by $\tilde{A}_j$. From this comparison we introduce the notion of \emph{zone} in \Definition{def:zone}, that we use to adjust the threshold. In \Fig{fig:zone}, that zone is enclosed by dashed lines and shaded in gray.
\begin{definition}{Zone ($Z$)}: Area around the expectation of the number of accepted \candidates inside which the threshold $\gamma_j$ is identical to that of \algo. It is defined between the two curves $\hat{\mu}_j +w_j$ and $\hat{\mu}_j - w_j$ where $w_j = w(j)$ is the function that defines the zone's thickness at step $j$. $Z = \int_0^n dx 2w(x)$.
\label{def:zone}
\end{definition}
\vspace{-4mm}
The threshold of this adjusted algorithm \emph{\algoNewfull} (\algoNew) is defined as:
\begin{equation}
\hat{\tau}^j(c) := 
\begin{cases}
\Xref{b}{c}, & \text{in the zone $Z$},\\
\Xref{m+D^+_{j-1}}{j} &  \text{below the zone $Z$},\\
\Xref{m-D^-_{j-1}}{j} &  \text{above the zone $Z$},
\end{cases}
\label{eq:adj_threshold}
\end{equation}
where $\Xref{i}{j} \in (\Sobold, \S{1},...,\S{j})$ is the score of the $i$-th best seen out of the \refset and up to the $j$-th \candidate \st $\Xref{1}{j} > ... > \Xref{b}{j}$, 
and $m\leq b$ is \, \st \, $\Xref{m}{j} = \Xref{b}{c}$. The $D^+_{j-1}$ and $D^-_{j-1}$ functions define how the threshold will change provided that a point $(j, \tilde{A_j})$ is outside the zone $Z$. More precisely, when that point lies in the zone, the threshold is constant and equal to $\Xref{b}{c}$.  When it is below (resp., above) the zone, then the threshold for the next \candidate is reduced (resp., increased) by $\lfloor D^+_j \rfloor$ (resp., $\lfloor D^-_j \rfloor$) positions from the former, as many times as needed until the point is inside the zone again, and the threshold goes back to the original one (\ie $\Xref{b}{c}$) for the next \candidate. Finally, $d_j = D^+_{j} - D^+_{j-1}=D^-_{j} - D^-_{j-1}$\, is the increment in the position each time a point has been above (resp. below) the zone in a row.
For simplicity, we assume that an optimal cutoff value $c^*$ for \algo is also an optimal cutoff value for \algoNew. The tuning of parameters $w_j$ and $d_j$ is done empirically, and the relevant ones are used in the simulations (see \Sec{sec:mssp}).

\section{A multi-round extension}\label{sec:mssp}

\subsection{General setting and assumptions}

In this section we build upon the \WSSP that was described thoroughly in previous sections, and introduce the \emph{\MSSPfull} (\MSSP). The process takes place in multiple successive rounds such that the output of a given round constitutes the input of the following one. The \emph{environment} of the problem is set to be on a large population $\mathcal{C}$ of job-seekers, \ie candidates. Essentially, each round constitutes a separate \WSSP (see \Sec{sec:ssp}) on a \emph{sample} of \candidates. 

\inlinetitle{Assumptions}{.} The \MSSP requires further assumptions: i)~the environment is considered to be fixed during each \WSSP round, however changes may occur between any two rounds regarding the \referents availability since any \referent can resign, and
ii)~sample are obtained by a random picking of \candidates in the population. The process may have an arbitrary number of \WSSP rounds. Therefore, the challenge for the \DM is to improve, or at least adapt, the personnel in the course of the multi-round process: at the end of any round that is to have selected the $\budget$-best items she could have chosen under the above assumptions and while respecting all the management constraints described for a single round in \Sec{sec:ssp}. We use the notations introduced in \Sec{sec:ssp} and add a subscript $k$ at each variable to refer to a precise round $k$, for instance $n_k$ is the number of \candidates at round $k$. 

\subsection{Implementing \algo in a \MSSP}	

In the previous section we created two algorithms (\algofull, \algoNew) that aim at selecting good \candidates in a single-round horizon. In this section, we intend to plug these algorithms in the multi-round setting (\MSSP) in order to iteratively improve the \DM's selection. For the simulations of this section we use the following parametrization. Firstly, each multi-round simulation considers a population of $|\mathcal{C}|=1000$ items and for all rounds we set the number of \candidates to $n=100$. Secondly, the resignation probability $\text{P}_{\nres,k} = \pres 
$ is considered to be known in advance by the \DM, and is kept constant for every round $k$ and equal for all \referents. 

\inlinetitle{Cutoff-choice and resignations}{.}
%
\begin{figure}[t!]
\centering
\hspace{-5mm}
\subfigure[$\pres = 0$]{\includegraphics[width = 0.35\linewidth]{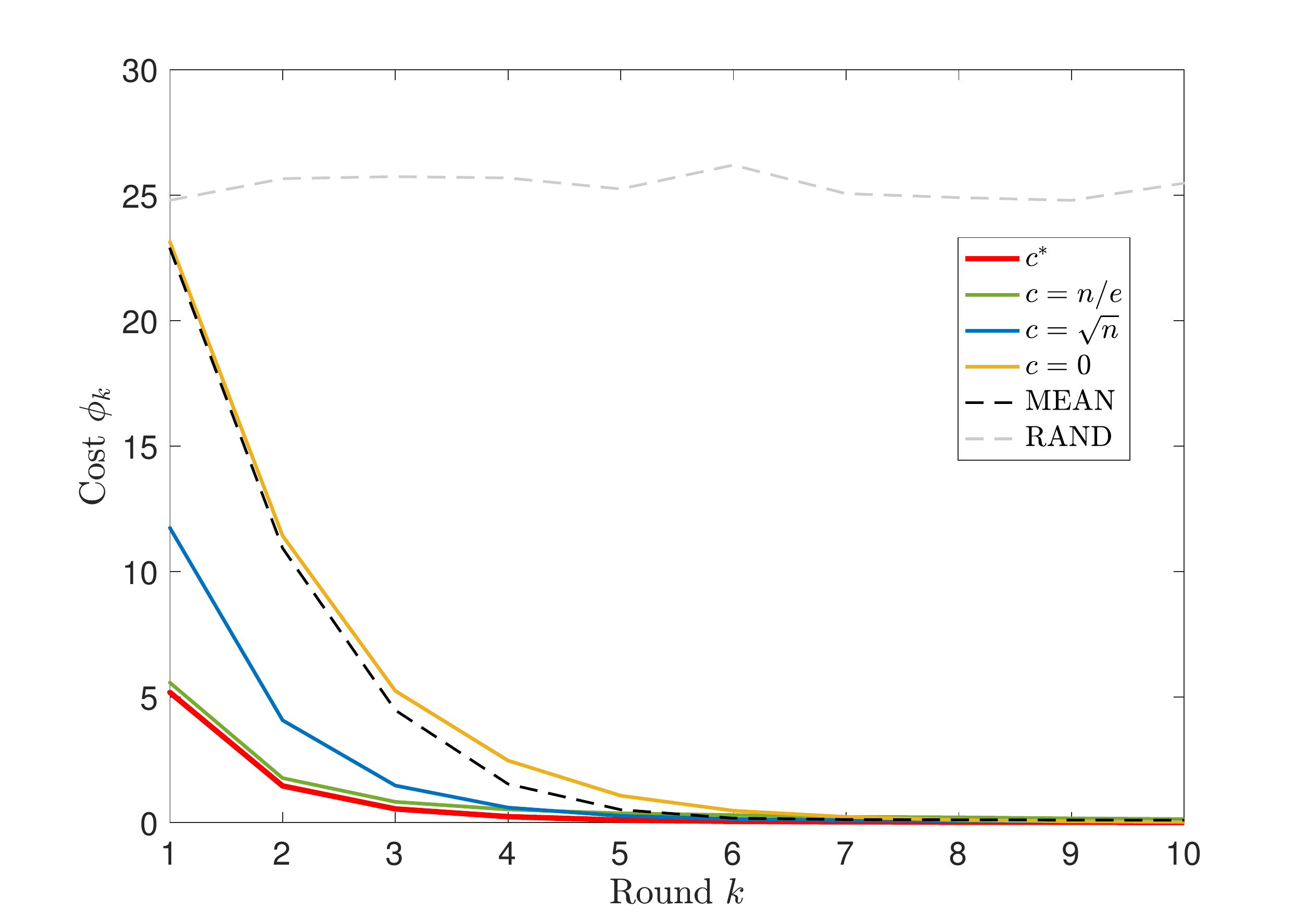}}%
\hspace{-3mm}
\subfigure[$\pres = 0.5$]{
\clipbox{18 0 0 0}{
\includegraphics[width = 0.35\linewidth]{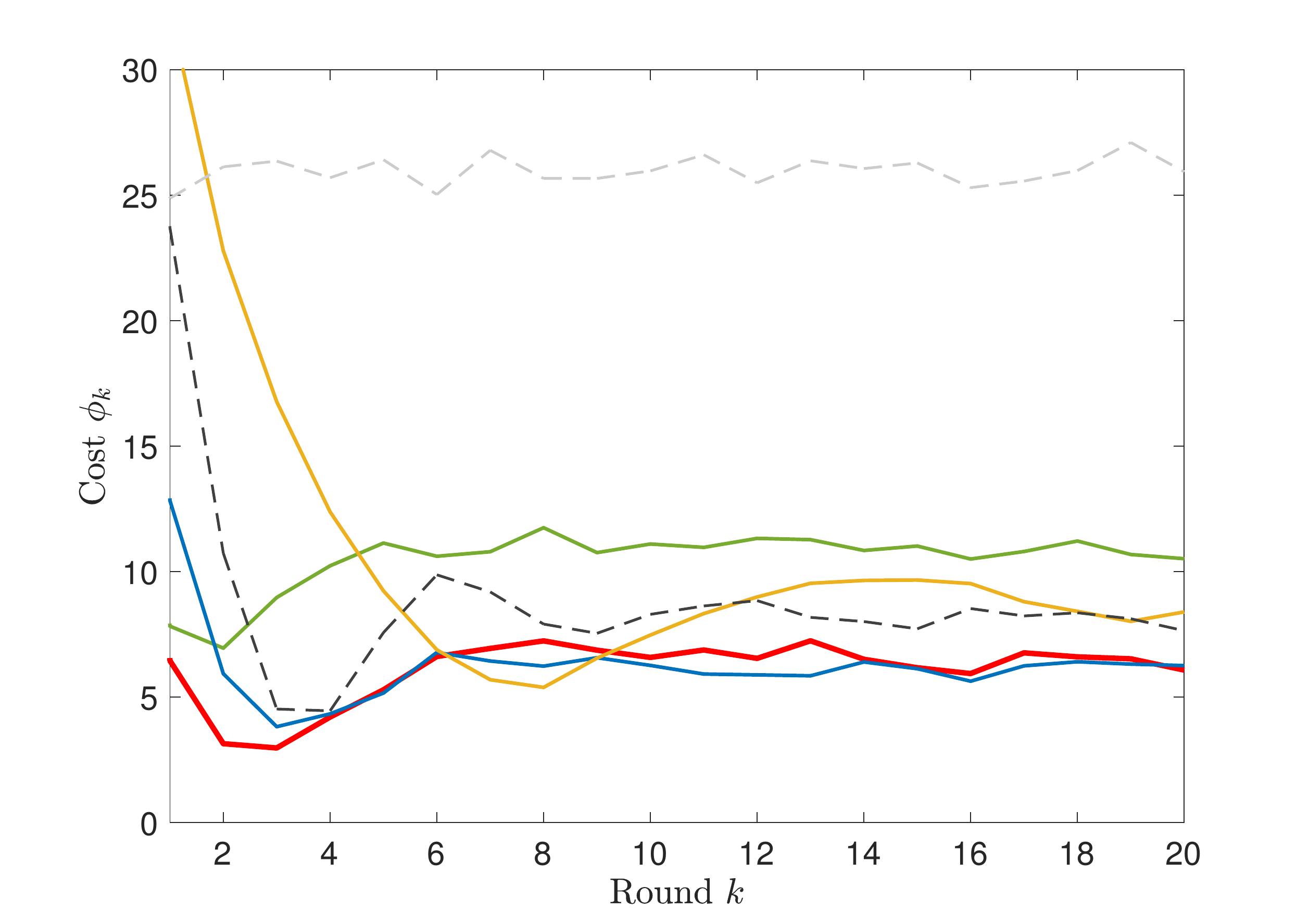}} }%
\hspace{-5mm}
\subfigure[$\pres = 1$]{
\clipbox{14 0 0 0}{\includegraphics[width = 0.35\linewidth]{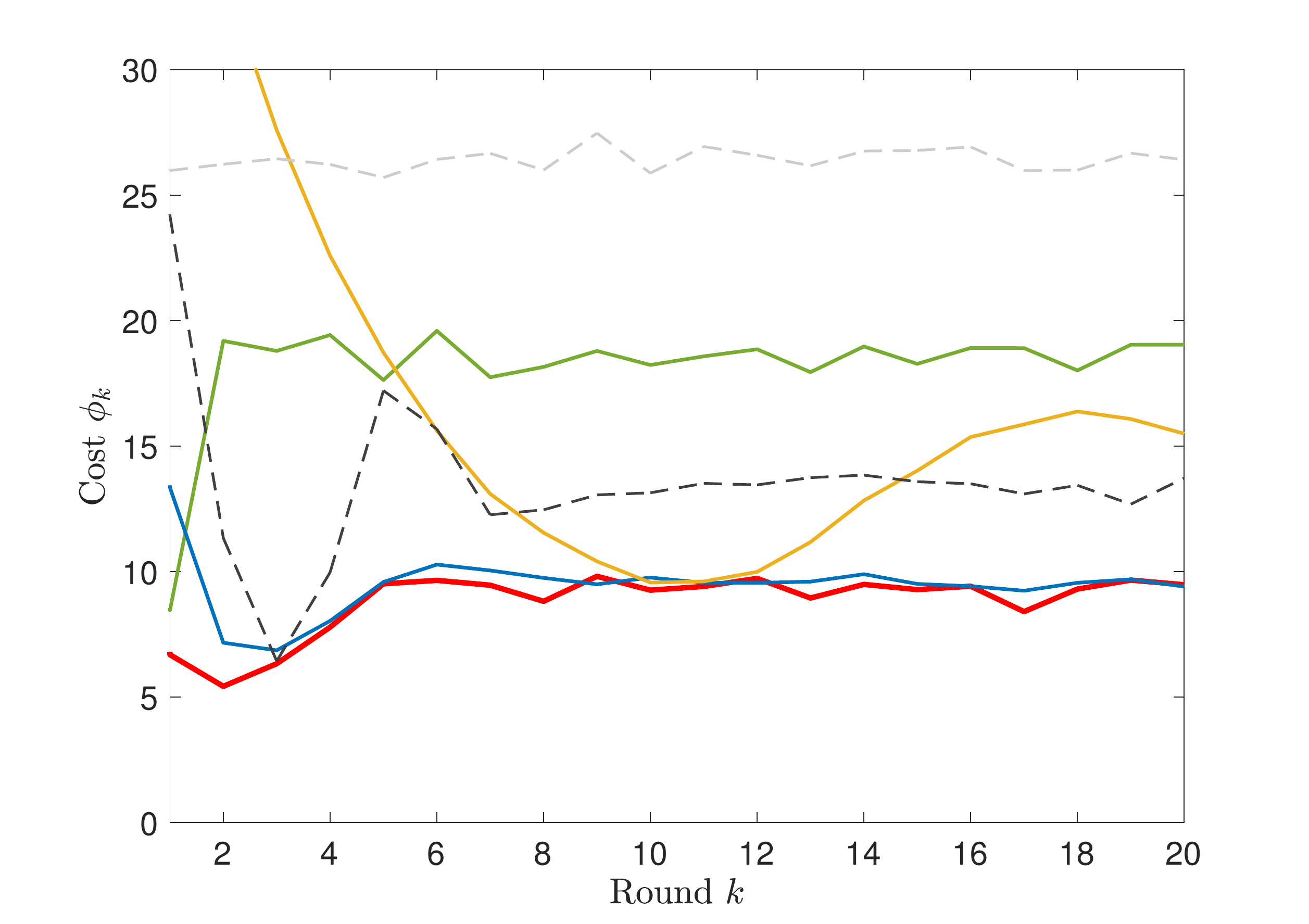}}}
\caption{\captionSize Average \regret ${\rsymb}_k$ \wrt the round number $k$ in the stationary case, for $n=100$ \candidates. The number of jobs is $b=5$.
 The dashed lines are not \algo strategies: MEAN accepts a \candidate if its score is above the mean of the current \referents and RAND accepts a \candidate if its score is above a randomly computed threshold.}
\label{fig:rSSP}
\end{figure}
\begin{figure}[t]
\centering
\hspace{-5mm}
\subfigure
{\includegraphics[width = 0.36\linewidth]{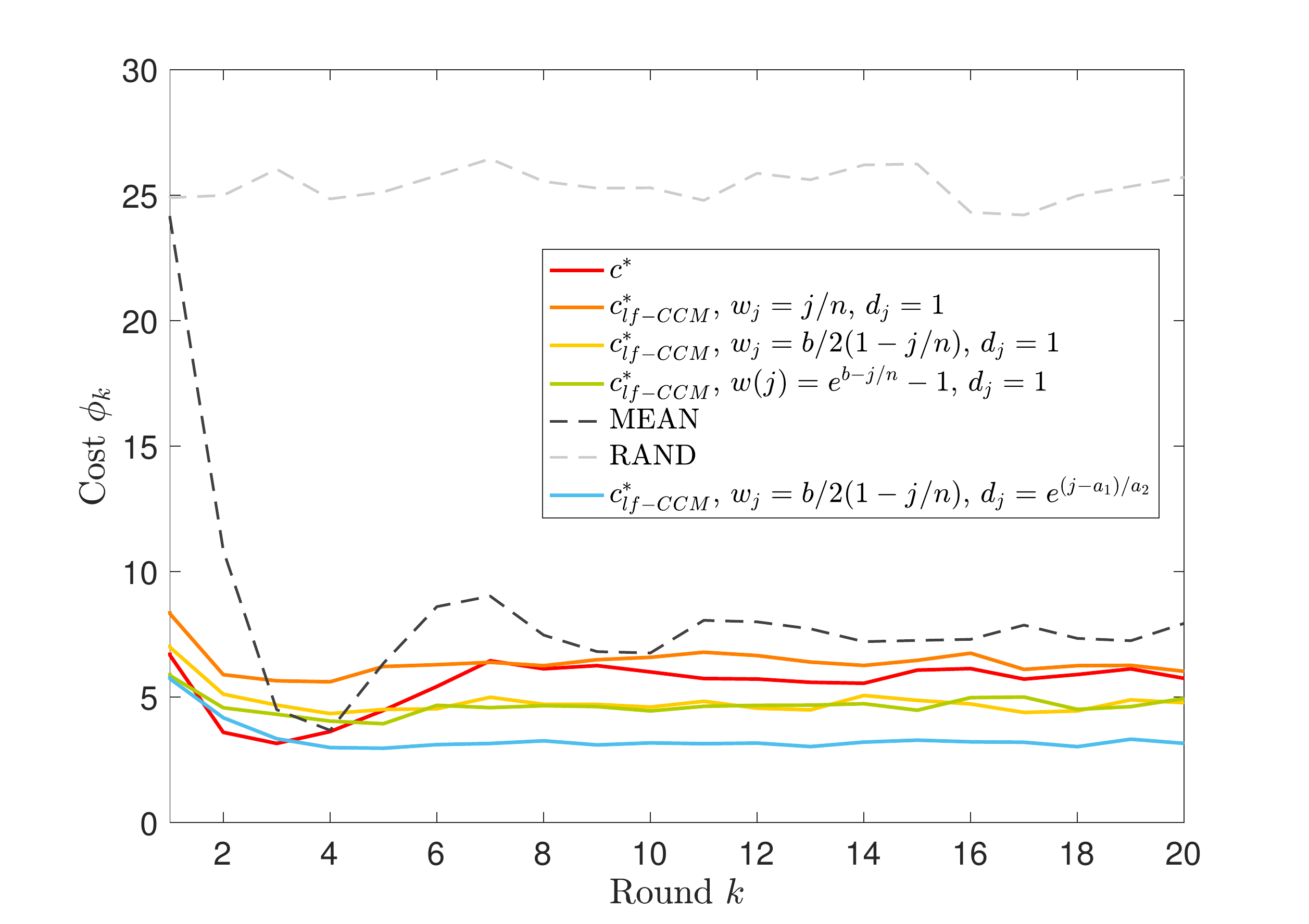}}
\hspace{-5mm}
\subfigure{ 
\clipbox{15 0 0 0}
{\includegraphics[width = 0.36\linewidth]{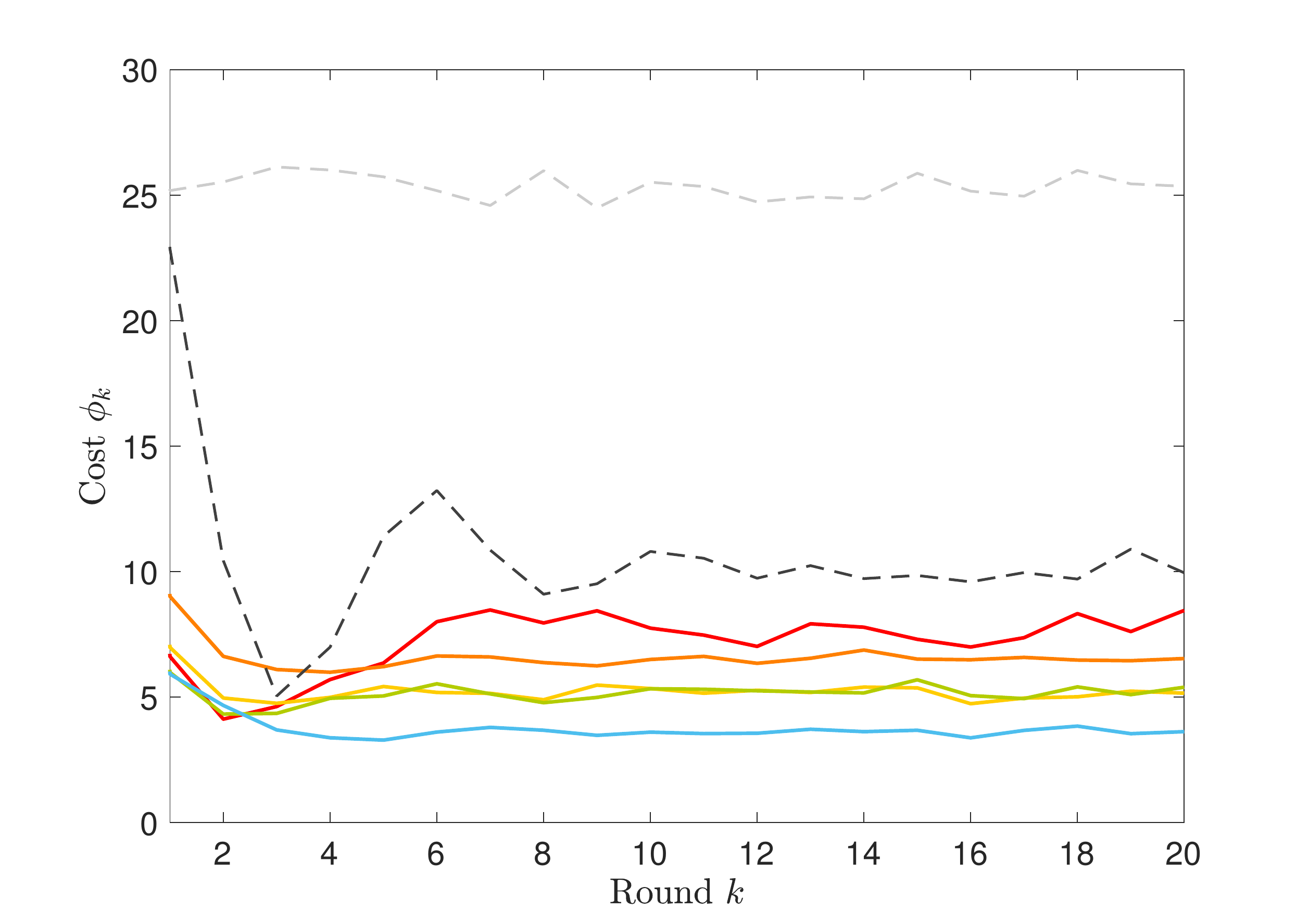}} }
\hspace{-5mm}
\subfigure{
\clipbox{15 0 0 0}
{\includegraphics[width = 0.36\linewidth]{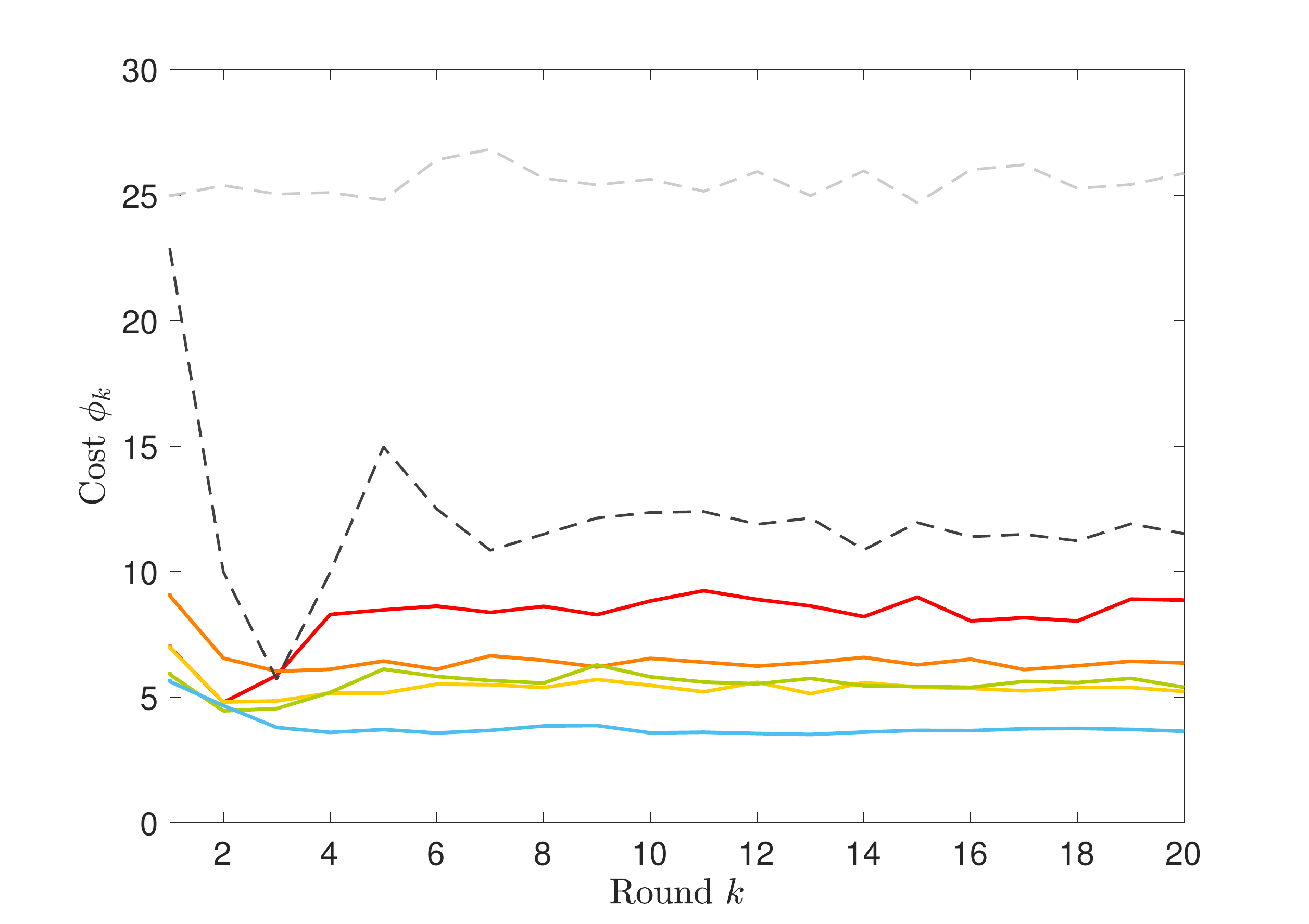}} }%
\vspace{-4mm}\\
\hspace{-5.5mm}
\subfigure[$\pres = 0.5$]{
\includegraphics[width = 0.36\linewidth]{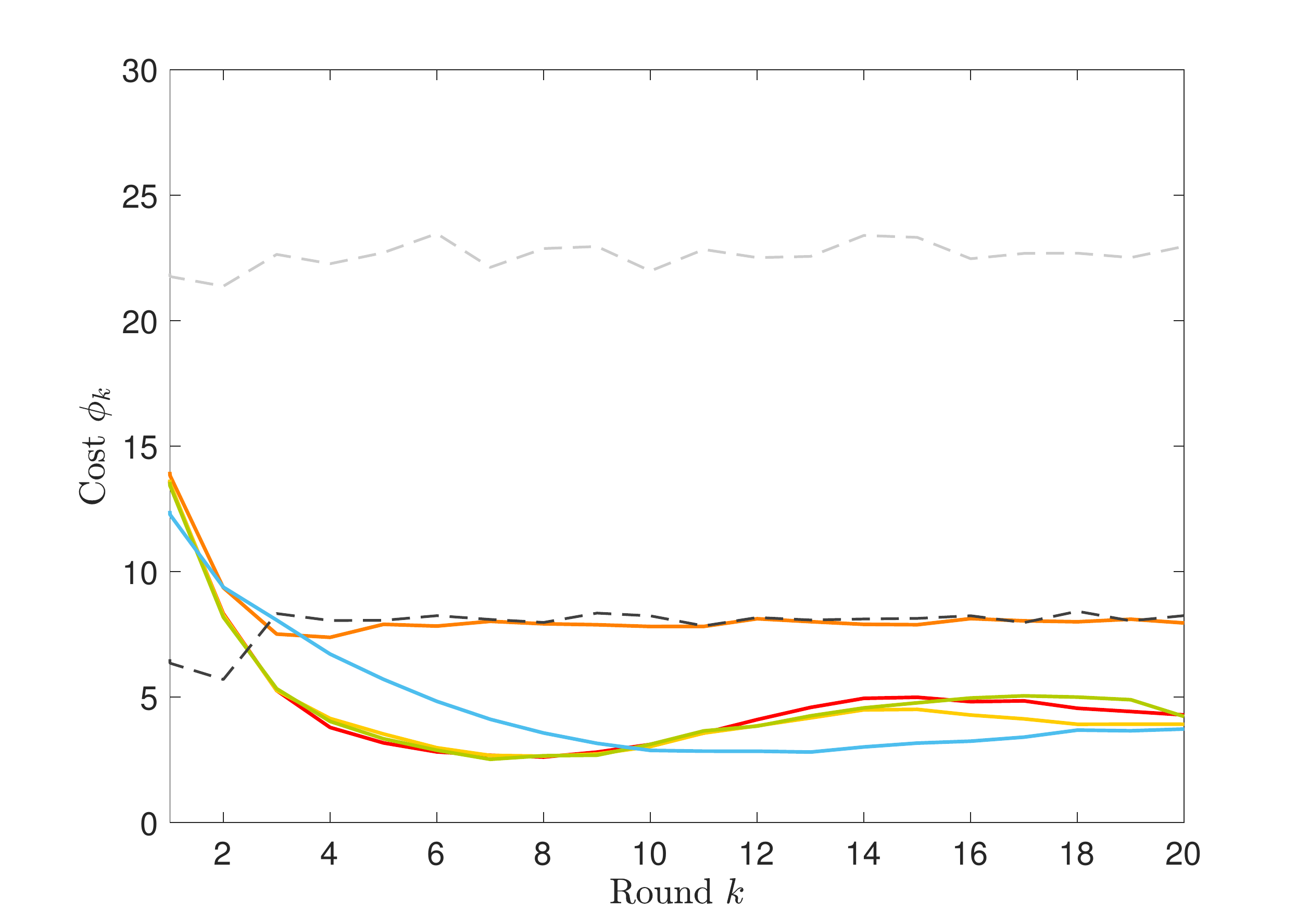}} 
\hspace{-5mm}
\subfigure[$\pres = 0.8$]{
\clipbox{15 0 2 0}
{\includegraphics[width = 0.36\linewidth]{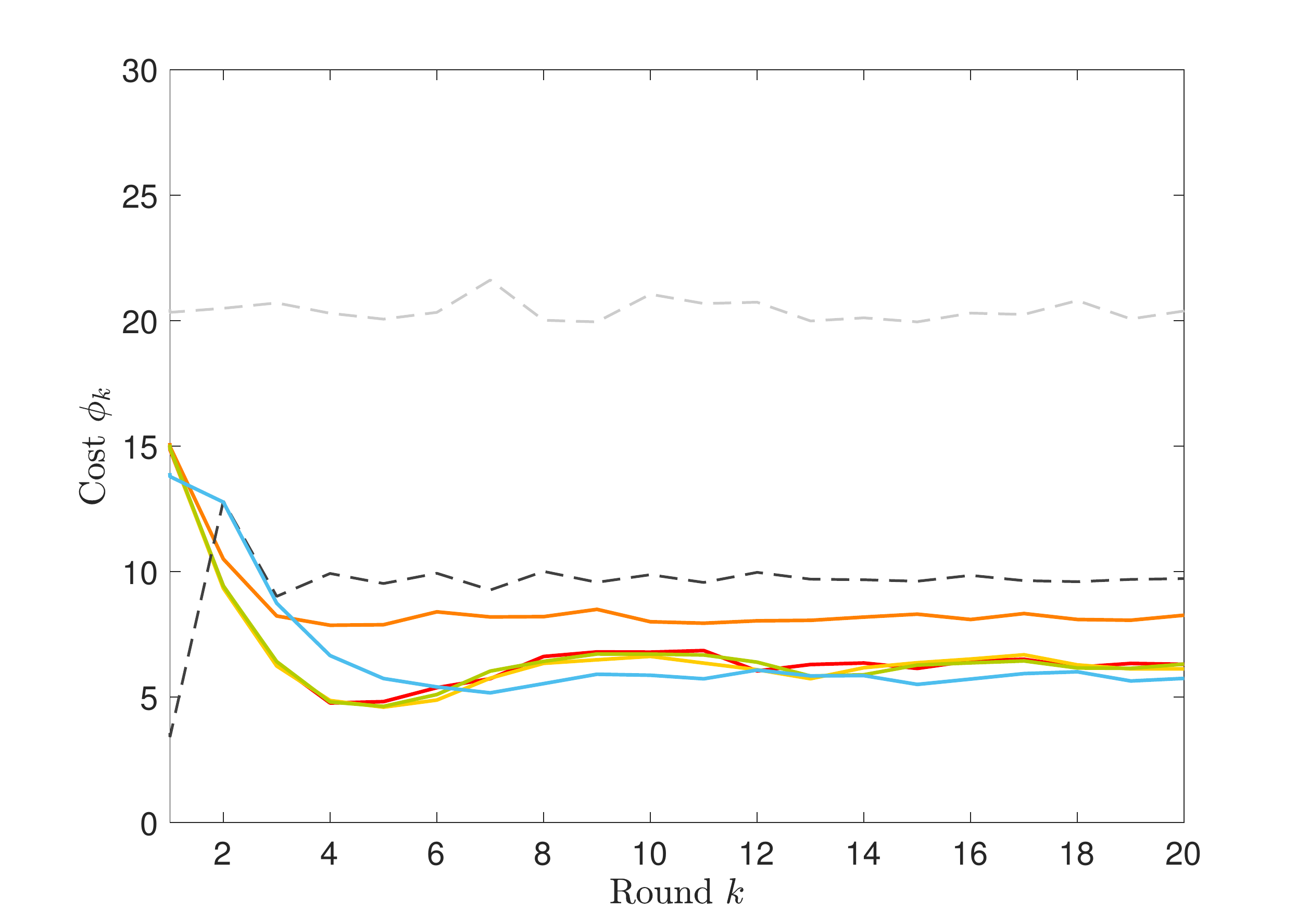}} }
\hspace{-5mm}
\subfigure[$\pres = 1$]{
\clipbox{15 0 0 0}
{\includegraphics[width = 0.36\linewidth]{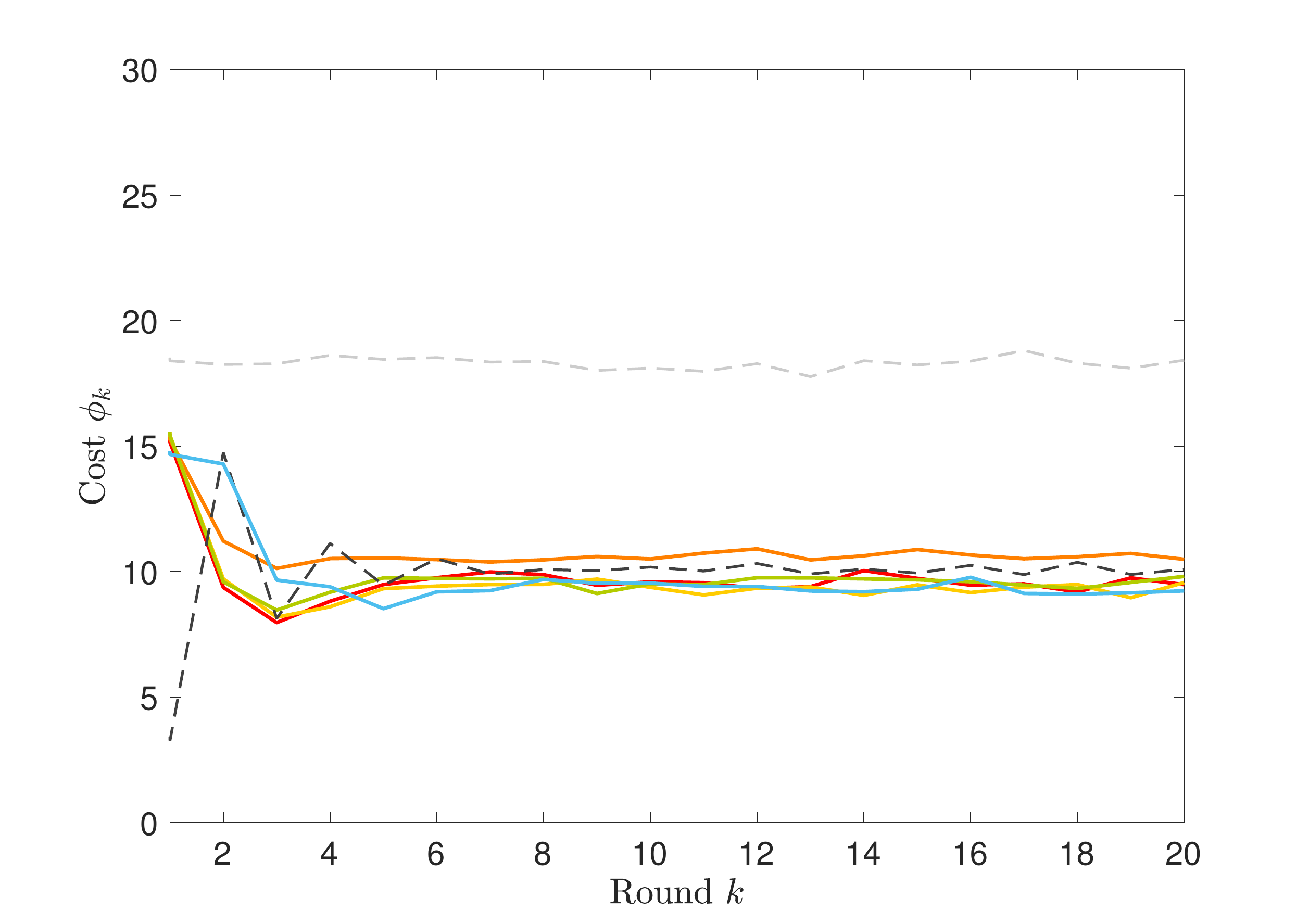}} }%
\caption{\captionSize Average \regret ${\rsymb}_k$ \wrt the round number $k$ in the stationary case, for $n=100$ \candidates. The red curve uses \algo when the orange yellow green and blue ones use \algoNew. The parameters for the blue curve are \st $v_1=82$ and $v_2 = 14$.  The dashed lines are not \algo strategies: MEAN accepts a \candidate if its score is above the mean of the current \referents and RAND accepts a \candidate if its score is above a randomly computed threshold. Top line $b=5$ and bottom line $b=50$.}
\label{fig:regret_different_stddev}
\end{figure}
\Fig{fig:rSSP}, displays the average \regret ${\rsymb}_k$ \wrt the round number $k$ for different resignation probabilities. We first observe that, regardless the resignation probability, our proposed cutoff $c^*$ (red curves) outperforms other alternatives originating from the general \SSP literature, or heuristics such as the case $c=0$. 
As presented, MSSP allows for \referents to resign their job at the beginning of a round, with probability $\pres$. Notice that, the cutoff $c = n/e$ is a decent alternative to $c=c^*$ when $\pres = 0$ (see \Fig{fig:rSSP}(a)), although failing at reducing the \regret when $\pres = 1$ (see \Fig{fig:rSSP}(c)). Large number of resignations can occur when the environment changes abruptly (\eg company's future, changes in the job market, \etc), or when the time-interval between two subsequent rounds is very long and more \referents may happen to resign. 

Another observation on this scenario is that \algo seems to struggle to make the \regret converge towards zero, and as stated in \Sec{sec:adjusted_algorithm} this effect is a consequence of being forced to select the last \candidate(s) in order to assign all vacant jobs (\ie failure), hence the \algoNewfull. A comparison of \algo and \algoNew can be found in \Fig{fig:regret_different_stddev} and illustrates the fact that \algoNew is more efficient at improving the selection through rounds than \algo, although it requires more adaptation from the \DM.

\section{Conclusion}\label{sec:conclusions}

In this paper we introduced the \WSSPfull (\WSSP), where a \DM has at hand a set of \referents, some of which still available, randomly incoming one-by-one. Following the well-known \SP, we developed a \emph{cutoff-based} strategy, the \algofull (and a tuned version \algoNewfull), composed of a learning phase and a selection phase. The optimal length of the former according to the number of initially empty jobs is an intriguing question for which we brought interesting and not always straight to see results. The rank-based \regret function that we used enables our algorithm to be efficient for arbitrary \candidate \scores. We approximate analytically this objective function by deriving main parameters'\ expectations in closed-form (\eg the acceptance threshold, the number of accepted \candidates, the \regret, \etc). 

In the second part of the paper we implemented our algorithm \algo in a multi-round framework (\MSSP). That process is motivated by the needs of real-world recruitment processes that are constantly trying to improve the personnel of an organization or a company.

The conducted simulations are consistent with our analytical work, and demonstrated that \algo is efficient in reducing the \regret at the course of the multi-round process while being robust to scores, resignations or number of jobs changes. Moreover, our experiments showed that our proposed optimal cutoff $c^*$ compares favorably against various cutoff values presented in literature for other sequential selection settings.

In our future work, we plan on adopting and testing \algo for \MSSP in various applications. In addition, the multi-round setting creates plenty of room for developing statistical learning methods aiming to learn efficiently \candidates scores, were they to come from a given distribution. 

\section*{Appendix - Technical proofs}

\proof{Proof of \Proposition{prop:gamma_0_res}.}
\Eq{eq:gamma_0} derives from the definition of the quality in \Definition{def:quality}, and uses the fact that $\Exp{\Xo{i}}=\frac{ \Exp{\Xo{b}} }{n}i= \frac{\gamma_0}{n}i,\, \ \forall i \leq b$. 

The best available \referent, \ie with rank $\Xavail{1}$ is therefore expected to have a rank at best $\gamma_0/\budget$ and at worst $\gamma_0 ({\nres}+1)/\budget$. 
He has an expected rank of $\gamma_0/\budget$ iff the available item(s) are any of the $\budget-1$ below him in the ranking, \ie with a probability $\binom{\budget-1}{{\nres}}/\binom{\budget}{{\nres}}$. Then, he has an expected rank of $2\gamma_0/\budget$ iff the best \referent resigned and the ${\nres}-1$ other unavailable \referents are any of the $\budget-2$ below him in the ranking, \ie with probability $\binom{\budget-2}{{\nres}-1}/\binom{\budget}{{\nres}}$. 
Finally: 
\begin{align}
\Exp{\Xavail{1}} &= \sum_{i=1}^{{\nres}+1} \Prob\left(\Xavail{1} = \Exp{\Xo{i}}\right) \Exp{\Xo{i}} 
= \sum_{i=1}^{{\nres}+1}\frac{\binom{\budget-i}{{\nres}+1-i}}{\binom{\budget}{{\nres}}} \frac{i\gamma_0}{\budget}
= \frac{\gamma_0}{\budget\binom{\budget}{{\nres}}} \sum_{i=1}^{{\nres}+1} \binom{\budget-i}{{\nres}+1-i}i\,,
\intertext{from the multiset relation $\sum_{i=0}^n \binom{m+i-1}{i} = \binom{n+m}{n}$ we obtain:}
\Exp{\Xavail{1}} &= \frac{\gamma_0}{\budget\binom{\budget}{{\nres}}} \bigg( ({\nres}+1) \binom{\budget+1}{{\nres}+1}-(\budget-{\nres}) \binom{\budget+1}{{\nres}} \bigg)= \frac{\gamma_0(\budget+1)}{\budget(\budget-{\nres}+1)}.
\end{align}
Using $\Exp{\Xavail{l}} = \Exp{\Xavail{1}} l$, $\forall l \in 
\INTerval{1}{\budget-{\nres}}$, we obtain $\Exp{\Xavail{l}} = \frac{\gamma_0(\budget+1)l}{\budget(\budget-{\nres}+1)}$.
\endproof{}
\proof{Proof of \Proposition{prop:r_opt}.}
We begin by deriving the variable $\eta$ that gives the expected number of \referents that belong to the $b$-best, \ie $\eta = \Exp{\sum_{i=1}^{\budget} \Ind{\Xavail{i} \leq \budget}}$: 
\begin{align}
\eta &= \frac{{\nres}}{\budget} \Exp{\sum_{l=1}^{\budget} \Ind{\Xo{l} \leq \budget}}
 = \frac{{\nres}}{\budget} \sum_{l=1}^{\budget} \Prob(\Xo{l} \leq \budget)
 = \frac{{\nres}}{\budget} \sum_{l=1}^{\budget} \Ind{\frac{\gamma_0l}{\budget} \leq \budget}
 = \frac{{\nres}}{\budget} \sum_{l=1}^{\budget^2/\gamma_0}\!1\ \ \Leftrightarrow\ \  
\eta = \frac{\nres\,\budget}{\gamma_0}.
\end{align}
In $(\Xobold, \Xbold) \in \rankset{n+b}$, \candidates and \refset are ranked jointly, regardless if the \referents resigned or not. The optimal \regret is defined as the average sum of the $\budget$-best available ranks. If one of the unavailable \referents is among the $\budget$-best, his rank is replaced by the next best available rank (same for multiple unavailable \referents), which increases the expected offline \regret. Formally:
\begin{align}
\Exp{\coff} &=  \frac{\budget-\eta}{\budget}\sum_{m=1}^\budget m + \sum_{m=\budget+1}^{\budget+\eta} m
= (\budget-\eta)\frac{\budget+1}{2} + \eta \budget + \frac{\eta(\eta+1)}{2}\ \ \Leftrightarrow\\
\Exp{\coff} &=  \frac{b(\budget+1+\eta+\eta^2/\budget)}{2}\ \ \Leftrightarrow\ \ 
\Exp{\coff} = \frac{b(\budget+1)}{2}+\frac{\nres b^2 (\gamma_0 + \nres)}{2\gamma_0^2}.  
\end{align}%
\endproof%
\proof{Proof of \Lemma{lem:g}.} 
Set $p_j = \Prob(\X{j} < \gamma_j)$,\,\, $Z_j \sim \text{Bernoulli}(j-c,p_j)$,\,\, and $\tilde{Z}_j = \sum_{i=c+1}^{j} Z_{i}$. Thus:
\begin{equation}
\Prob(\ajm < \budget) = \Prob\bigg(\text{min}\big(\sum_{i=c+1}^{j-1} Z_i, b\big) < b\bigg) = \Prob\bigg(\sum_{i=c+1}^{j-1} Z_i < b\bigg) = \Prob(\tilde{Z}_{j-1} < b).
\end{equation}
We have $\quality = 1/2$, hence $(\Xobold,\Xbold)$ is uniformly distributed in $\INTerval{1}{n+\budget}$. Therefore, $ p_j = \frac{\gamma_j-1}{n+b} $ and since $j \leq n$, $j\rightarrow \infty \Rightarrow n\rightarrow \infty$ thus, $\underset{j\rightarrow \infty}{\lim} \sum_{i=c+1}^j (\frac{\gamma_i-1}{n+b})^2 = 0$. Therefore $\underset{j\rightarrow \infty}{\lim} \sum_{i=c+1}^j p_i^2=0$, in other words, the more candidates there are, the smaller the probability for each of them to be accepted. Set $\smalloj =  \sum_{i=c+1}^j p_i^2$ and $\lambda_j = p_{c+1}+...+p_j$. From Le Cam's theorem we have $\sum_{m=0}^\infty |\Prob(\tilde{Z}_j=m) - \frac{\lambda_j^m e^{-\lambda_j}}{m!}| < 2 \smalloj$, and since $\underset{j\rightarrow \infty}{\lim} \smalloj =0$, $\tilde{Z}_j $ tends a Poisson distribution with parameter $\lambda_j$. Its cumulative distribution is therefore, $\Prob(\tilde{Z}_j\leq x) = \Prob(\tilde{A}_j\leq x) = e^{-\lambda_{j-1}} \sum_{i=0}^{x}\frac{  \lambda_{j-1}^i}{i!} + o(\smalloj)$.
\endproof
\proof{Proof of \Theorem{th:expected_cost}.}We handle each bullet point separately:

\begin{itemize}[itemsep=0pt,leftmargin=*]
\item First, we investigate the rank-based expected threshold to beat for the first \candidate incoming just after the learning phase, $\gamma := \Exp{\Xref{b}{c}}$.
The proof is done by backward induction. We first consider the case where the number of rejected \candidates $c$ is \st $c=n$; the \updref is composed of the $\budget$-best items of $(\Xobold, \Xbold)$ since every \candidate has been rejected and their scores are stored in the \updref. Thus $\gamma=\budget$. Let us go one step ahead and consider the case where $c=n-1$, which implies that $\gamma=\budget$ if the \candidate that has not been examined is not among the $\budget$-best items, and $\budget+1$ if he is. Hence, $\gamma(c=n-1) = \budget \frac{c}{\budget+c} + (\budget+1)\frac{\budget}{\budget+c}$. By recursion, we get:
\begin{align}
\gamma(c) & = \sum_{m=0}^{n-c} \binom{n-c}{m} (\frac{c}{\budget+c})^{n-c-m}(\frac{\budget}{\budget+c})^{m}(\budget+m) = \frac{1}{(\budget+c)^{n-c}}\sum_{m=0}^{n-c} \binom{n-c}{m} c^{n-c-m}\budget^{m}(\budget+m)\\
&\Leftrightarrow\ \ \gamma(c) = \frac{\budget(n+\budget)}{\budget+c}.
\end{align}
When $j >c$, after multiple repetitions of the selection, each acceptance threshold is replaced by its expectation, in particular $\Xref{b}{c}$ tends towards its expectation $\gamma := \Exp{\Xref{b}{c}}$. Hence $\delta = r+ \sum_{j=1}^c \Ind{\X{j} < \Xref{b}{c}}$ tends to $\Exp{r+ \sum_{j=1}^c \Ind{\X{j} < \gamma}}$, \ie $\Delta := \Exp{\delta} = r + \sum_{j=1}^c \Prob(\X{j} < \gamma) = r + c\frac{\gamma-1}{n+b}$. Then, the evolving threshold becomes $\gamma_j = \gamma\Exp{\Ind{\ajm < \Delta}} + \Exp{\Xavail{\budget - \ajm}\Ind{\ajm \geq \Delta}}$. In order to use the fact that $\Exp{\Xavail{l}} = \Exp{\Xavail{1}} l$, $\forall l \in \INTerval{1}{\budget-{\nres}}$, in the proof we approximate $\Xavail{\budget - \ajm}$ by $\Xavail{\budget - \Exp{\ajm}}$ by considering that  $\aj$ has a small variance, which is given by $\smallo$. Therefore:
\begin{align}
\gamma_j & = \gamma \Exp{\Ind{\ajm < \Delta}} + \Exp{\Xavail{\budget - \Exp{\ajm}}}\Exp{\Ind{\ajm \geq \Delta}} +o(\smallo)\\
&= \gamma \Prob(\ajm < \Delta) + \frac{\gamma_0(\budget+1)}{\budget(\budget-{\nres}+1)}(\budget-\Exp{\ajm})\Prob(\ajm \geq \delta)+o(\smallo) 
\intertext{We have $\Exp{\aj} = \sum_{i=c+1}^j \Exp{\A{i}}=  \sum_{i=c+1}^j \Prob(\A{i}=1)=\sum_{i=c+1}^j \Prob(\X{i}<\gamma_i)\Prob(\tilde{A}_{i-1}<b)$; hence:}
\gamma_j &= \gamma g_j(\Delta) + \frac{\gamma_0(\budget+1)}{\budget(\budget-{\nres}+1)}\left(\budget-\sum_{i=1}^{j-1}\frac{\gamma_i-1}{n+b}g_i(\budget)\right)(1-g_j(\Delta)) +o(\smallo).
\end{align}
where $g_j(x) := \Prob(\ajm < x)$ is computed using \Lemma{lem:g}. 
 
\item Since $\tau_j$ tends to $\gamma_j$, we get $A_j =\Ind{j >c} \Ind{\ajm < b}\Ind{\X{j} < \gamma_j}$, hence:
\begin{align} 
\Exp{\an} &:= \sum_{j=1}^n \Exp{\A{j}} = \sum_{j=1}^n \Prob(\A{j}=1)= \sum_{j=1}^n \Prob(\X{j} <\gamma_j)\Prob(\ajm < b)= \sum_{j=1}^n \pjnd g_j(b).
\end{align}
\item Recall the definition of the \regret $\rsymb :=  \Xobold\scalar \Aobold{n}+ \Xbold\scalar\Abold -\coff$. Set $\rsymb_1 =  \Xobold\scalar \Aobold{n}$ and $\rsymb_2 = \Xbold\scalar\Abold$ that give respectively the \refset and the candidates contribution to the \regret. We start with the candidates, $\rsymb_2 = \sum_{j=1}^{n} \X{j} \A{j}$. Its expectation is given by $\Exp{\rsymb_2} =  \oneoverb \Exp{\sum_{j=1}^{n} \X{j} \A{j}}$. We use the fact that $\A{j}=0, \, \ \forall j \leq c$:
\begin{align} 
\Exp{\rsymb_2} &= \oneoverb \sum_{j = c+1}^{n} \sum_{m=1}^{n+\budget} \sum_{a= \{0,1\}}\Prob(\X{j} =m, \A{j}=a)am
=\oneoverb\sum_{j = c+1}^{n} \sum_{m=1}^{n+\budget}  \Prob(\A{j}=1 \mid \X{j} =m) \Prob(\X{j} =m)m
\intertext{A \candidate with rank higher than the threshold $\gamma_j$ will never be accepted, hence:}
\Exp{\rsymb_2} &=\oneoverb\sum_{j = c+1}^{n} \sum_{m=1}^{\gamma_j-1}  \Prob(\A{j}=1 \mid \X{j} =m) \Prob(\X{j} =m)\,m.
\intertext{A \candidate with rank lower than the threshold is accepted if there were less than $\budget$ \candidates accepted before him. Moreover, we use the fact that $\Prob(\X{j} =m) = \frac{1}{n+b}$ to write:}
\Exp{\rsymb_2} &= \oneoverb \sum_{j = c+1}^{n} \sum_{m=1}^{\gamma_j-1}  \Prob(\ajm < b) \frac{m}{n+b}
=\oneoverb \sum_{j = c+1}^{n} \sum_{m=1}^{\gamma_j-1} g_j(\budget)\frac{m}{n+b}
= \frac{1}{n+b} \sum_{j = c+1}^{n} g_j(\budget) \frac{\gamma_j(\gamma_j-1)}{2}. \label{eq:proof_wofail}
\end{align}
Following up with the \refset contribution: 
$\rsymb_1 =  \sum_{l=1}^{\budget} \Xavail{l}\Ind{l \leq \budget - \an} =  \sum_{l=1}^{\budget- \an} \Xavail{l} $ is the regret associated with the available \referents that were not fired at the end of the selection. 
Its expectation is given by $\Exp{\rsymb_1} = \oneoverb \Exp{\sum_{l=1}^{\budget- \an} \Xavail{l}}$. We suppose that variables $\an$ and $\Xo{l}$ are independent $\forall l$, which is a reasonable assumption since we consider a \refset with medium quality, \ie medium average rank, and we use $\Exp{\Xavail{l}} = \frac{\gamma_0(\budget+1)l}{\budget(\budget-{\nres}+1)}$ (see \Proposition{prop:gamma_0_res}):
\begin{align}
\Exp{\rsymb_1} &= \oneoverb \sum_{l=1}^{\budget- \Exp{\an}} \frac{\gamma_0(\budget+1)l}{\budget(\budget-{\nres}+1)}
= \frac{\gamma_0(\budget+1)}{\budget(\budget-{\nres}+1)} \sum_{l=1}^{\budget- \Exp{\an}} l = 
\frac{\gamma_0(\budget+1)}{2\budget(\budget-{\nres}+1)} (\budget- \Exp{\an})(\budget+1 - \Exp{\an}). \label{eq:proof_pres}
\end{align}
\end{itemize}
\endproof

\proof{Proof \Proposition{prop:mu_hat}.} Set $\tilde{Z}_j = \sum_{i=c+1}^j Z_i$ where $Z_j \sim \text{Bernouilli}(j-c,p_j)$ and set $\lambda_j = \sum_{i=c+1}^j p_i,  \, \ \forall j$. We have $\Exp{\aj \mid \an \geq \nres} = \sum_{k=0}^{\budget} k\Prob(\aj = k \mid \an \geq \nres)$. Hence:
\begin{align*}
 \Exp{\aj \mid \an \geq \nres} &= \sum_{k=0}^{\budget} k\frac{\Prob \big(\text{min}(\tilde{Z}_j,b)= k , \text{min}(\tilde{Z}_n,b) \geq \nres)\big)}{\Prob(\an \geq {\nres})}\\
\Exp{\aj \mid \an \geq \nres} &= \frac{1}{\Prob(\an \geq {\nres})}  \sum_{k=0}^{\budget-1} k\Prob\bigg(\{\tilde{Z}_j = k\} \cap \{\tilde{Z}_n \geq {\nres}\} \bigg) + \frac{\budget}{\Prob(\an \geq {\nres})}  \Prob\bigg(\{\tilde{Z}_j \geq \budget \}  \cap \{ \tilde{Z}_n \geq {\nres}\}\bigg) \\
\Exp{\aj \mid \an \geq \nres} &= \frac{\Prob(\tilde{Z}_n \geq {\nres})}{\Prob(\an \geq {\nres})} \sum_{k=0}^{\budget-1} k\Prob(\tilde{Z}_j = k) + \frac{\budget\Prob(\tilde{Z}_j \geq b) }{\Prob(\an \geq {\nres})}\\
\Exp{\aj \mid \an \geq \nres} &= \sum_{k=0}^{\budget-1} k\Prob(\tilde{Z}_j = k) +  \frac{\budget(1-g_{j+1}(\budget))}{1-g_{n+1}({\nres})}.
\intertext{From Le Cam's theorem we have $\sum_{m=0}^\infty |\Prob(\tilde{Z}_j=m) - \frac{\lambda_j^m e^{-\lambda_j}}{m!}| < 2 \smalloj$, and since $\underset{j\rightarrow \infty}{\lim} \smalloj =0$, $\tilde{Z}_j $ tends a Poisson distribution with parameter $\lambda_j$, see proof of \Lemma{lem:g}, hence:}
\Exp{\aj \mid \an \geq \nres} &= \sum_{k=0}^{\budget-1} \frac{\lambda_j^k e^{-\lambda_j}}{(k-1)!} +  \frac{\budget(1-g_{j+1}(\budget))}{1-g_{n+1}({\nres})} + o(\smalloj)\\
\Exp{\aj \mid \an \geq \nres} &=  \lambda_j \sum_{k=0}^{\budget-2} \frac{\lambda_j^{k} e^{-\lambda_j}}{
k!} +  \frac{\budget(1-g_{j+1}(\budget))}{1-g_{n+1}({\nres})} + o(\smalloj) \\
\Exp{\aj \mid \an \geq \nres} &=  \lambda_j g_{j+1}(b-1) +  \frac{\budget(1-g_{j+1}(\budget))}{1-g_{n+1}({\nres})}.  
\end{align*}
\endproof

\bibliographystyle{acm} 
\bibliography{bibliography}
\end{document}